\documentclass[preprint, superscriptaddress]{revtex4-1}
\usepackage{graphicx}
\usepackage{amsmath}
\usepackage{float}
\usepackage{subfig}
\usepackage{natbib}
\begin{document}
\title{Effects of DC voltage on initiation of whirling motion of an electrostatically actuated nanowire oscillator}
\author{A. Bhushan}
\affiliation{Department of Mechanical Engineering, National Institute of Technology Patna, Patna - 800005, India.}
\author{M. M. Inamdar}
\thanks{Corresponding author}
\email[Email address: ]{minamdar@iitb.ac.in}
\affiliation{Department of Civil Engineering, Indian Institute of Technology Bombay, Mumbai - 400076, India.}
\author{D. N. Pawaskar}
\thanks{Corresponding author}
\email[Email address: ]{pawaskar@iitb.ac.in}
\affiliation{Department of Mechanical Engineering, Indian Institute of Technology Bombay, Mumbai - 400076, India.}
\begin{abstract}
Planar driven nanowire oscillators are susceptible to undergo whirling motion due to coupling between flexural planar and nonplanar modes of vibration. This investigation is concerned with planar to whirling motion transition in the oscillation of an electrostatically actuated nanowire, which is pre-deflected due to applied DC voltage. We have derived dynamical equations of motion using Euler-Bernoulli beam theory and Galerkin formulation as reduced order model of the governing coupled partial differential equations. The dynamical equations have been solved using second-order averaging method and the averaging solution has been validated by comparing with the numerical solution of the reduced order model. Further, planar to whirling motion transition has been investigated by studying qualitative changes in resonance curves with variation of electrostatic actuation. In this paper, we provide a simple analytical condition which takes into account the effects of DC voltage for prediction of initiation of whirling motion. Our main observation of the present investigation is that DC voltage can tune the initiation pattern of the whirling motion and change the qualitative nature of resonance curves.
\end{abstract}
\maketitle
\section{Introduction}
Investigations of nanoelectromechanical system (NEMS) oscillators have received considerable attention in recent years due to their potential applications as ultra-sensitive sensors and signal-processing devices \cite{eom2011, dai2009, rutherglen2009}. Nonlinear study of NEMS oscillators is the thrust of many concurrent investigations to enhance the understanding of dynamic behaviour of these miniaturised devices \cite{rasekh2010, kozinsky2006, ouakad2010}. Nonlinear study is also motivated by some research explorations which observed that nonlinear responses can also be utilised to enhance the performance of NEMS oscillators \cite{younis2009, buks2006, karabalin2011}. For example, nonlinear resonance curves can be utilised to design ultra-sensitive NEMS mass-sensors \cite{dai2009, younis2009}. Similarly, by operating the nano-oscillators in nonlinear regime, sensitivity of NEMS mass sensors can be increased beyond the limit imposed by thermomechanical noises on linear regime operation \cite{buks2006}. Moreover, bifurcation topology of nano-resonators has applications in signal-amplification and switching operations \cite{karabalin2011}. This paper is concerned with the investigation of nonlinear resonance behaviour of an electrostatically actuated doubly-clamped cylindrical beam which represents a typical nanowire.\\
\begin{figure*}
\begin{center}
\includegraphics[width=7.5cm]{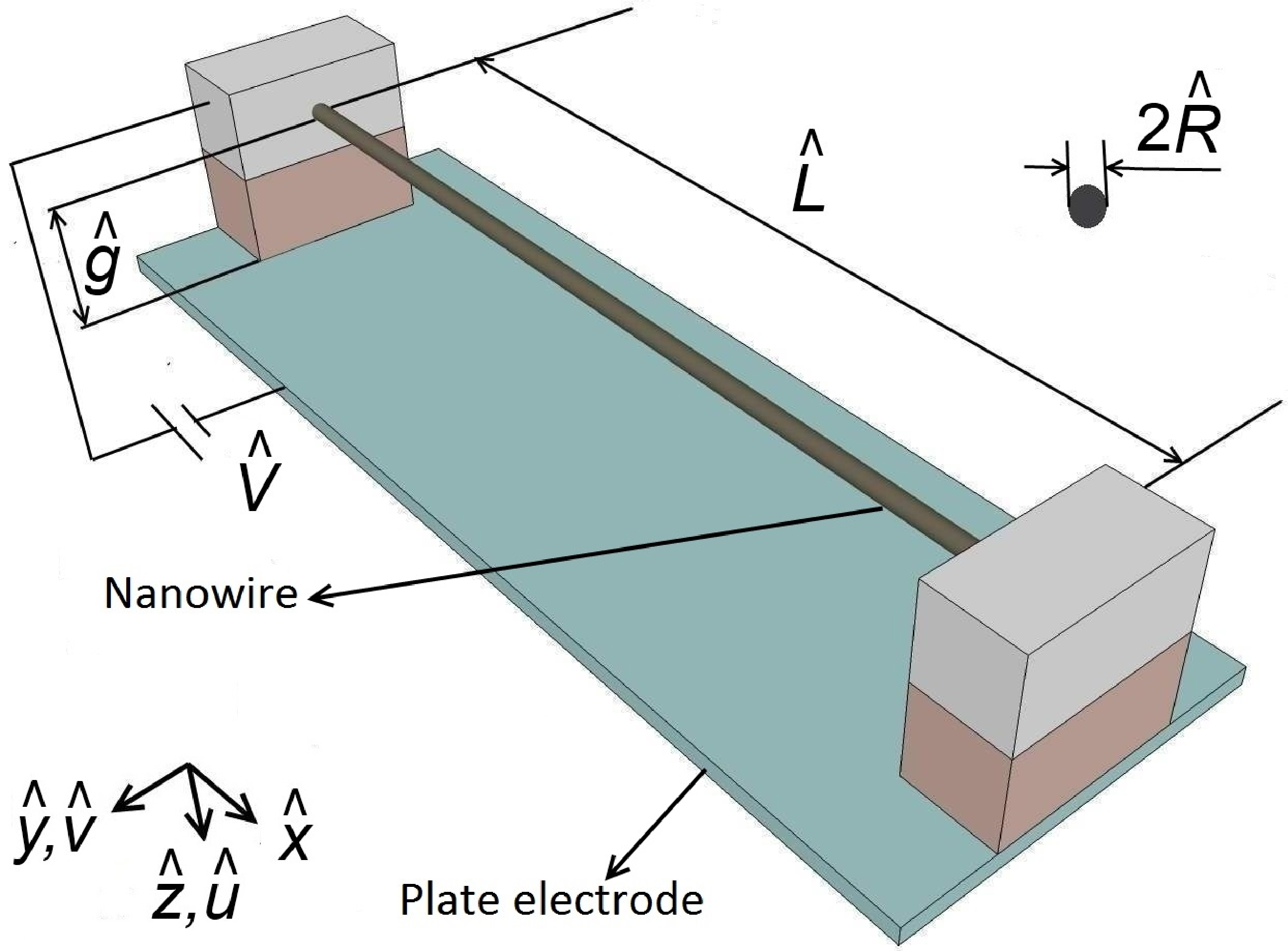}
\caption{Schematic diagram of an electrostatically actuated nanowire oscillator.}
\label{wfig.nanowire-model}
\end{center}
\end{figure*}
\indent In electrostatic actuation, a beam is placed parallel to a plate electrode and the oscillator is driven by applying bias DC and AC voltages between them. Electrostatically driven such nano-oscillator can undergo planar motion, where it oscillates in plane parallel to the axis of the nano-oscillator and perpendicular to the plate electrode (X-Z plane in Fig. \ref{wfig.nanowire-model}). Planar dynamics of nano-oscillators have earlier been investigated to study resonance behaviour \cite{rasekh2010, dequesnes2004}. Dequesnes et al. \cite{dequesnes2004} have investigated variation of first resonance frequency with respect to DC voltage of single-walled carbon nanotubes using a combination of molecular dynamic simulations and beam theory. On the other hand, Ouakad and Younis \cite{ouakad2010} have numerically investigated planar primary and secondary resonances of carbon nanotube oscillators using Galerkin based reduced order modelling technique. In another study on carbon nanotube oscillators, Rasekh et al. \cite{rasekh2010} have also numerically studied planar primary resonance behaviour using reduced order modelling. In an experimental investigation, Zhu et al. \cite{zhu2009} have examined zinc oxide nanowires as electromechanical oscillators for actuation and sensing purposes. Moreover, Solanki et al. \cite{solanki2010} have experimentally investigated nonlinear resonance behaviour of electrostatically actuated InAs nanowires.

\indent In addition to the literature of planar vibration of nano-oscillators, there also exist some research works where oscillatory components of a nano-oscillator exist in both perpendicular and parallel planes to the electrode plate, in other words, both planar (X-Z plane in Fig. \ref{wfig.nanowire-model}) and nonplanar (X-Y plane in Fig. \ref{wfig.nanowire-model}) directions respectively. For example, planar and nonplanar flexural modes of vibration of initially curved carbon nanotubes have been investigated by Ouakad and Younis \cite{ouakad2011}, where they have studied mode veering phenomenon and variation of natural frequencies with respect to magnitude of applied DC voltage. In an experimental investigation,  Abhilash et al. \cite{abhilash2012} have probed planar and nonplanar vibrations of InAs nanowires after electrostatically driving the nanowires through a side electrode that is placed along with the nanowires above the substrate. In another study, Gil-Santos et al. \cite{santos2010} have developed nanomechanical mass sensing technique by utilising resonance behaviour of planar and nonplanar modes of vibration of cantilever nanowires having imperfect cross-sectional geometry; they have used optical techniques to observe resonance peaks of the orthogonal modes of vibration.

\indent In nanowire and carbon nanotube oscillators, there is a nonlinear geometric coupling between flexural orthogonal planar and nonplanar modes of vibration which may interact through internal resonance \cite{conley2008, eichler2012}. Few earlier investigations have demonstrated that planar driven nanowire and nanotube oscillators may exhibit a change from planar to whirling motion at some critical magnitude of electrostatic actuation due to this nonlinear geometric coupling  \cite{conley2008, chen2010}; here whirling motion refers to the existence of oscillatory components in both planar and nonplanar directions. Specifically, Conley et al. \cite{conley2008} have investigated whirling dynamics of electrostatically actuated nano-oscillators using reduced order modelling for providing theoretical understanding of experiments of Sazonova et al. \cite{sazonova2004}. The authors have modelled a carbon nanotube oscillator, after neglecting the effects of initial deflection due to DC voltage, and provided an analytical criterion for the initiation of whirling motion using first-order averaging method. The same group, in a later investigation, also performed numerical analysis to study the role of parametric excitation terms introduced by DC voltage on resonance curves of nanotube oscillators \cite{conley2010}.  In another study, Chen et al. \cite{chen2010} have numerically investigated chaotic responses in whirling dynamics of nanowire oscillators by numerically solving governing partial differential equations of motion using a combination of finite element and Runge-Kutta methods. Additionally, Eichler et al. \cite{eichler2012} have experimentally studied the interaction among various orthogonal planar and nonplanar modes of vibration of nanotube oscillators. Apart from nano-oscillators, sinusoidally forced strings and suspended cables are other mechanical systems where whirling motion has been studied earlier \cite{arafat2003, pai1992, johnson1989, lee1995, abe2010}.\\
\indent Although the aforementioned studies on whirling dynamics of planar driven electrostatically actuated nano-oscillators provide insights into planar to whirling motion transition, much still remains to be understood. For instance, Conley et al. \cite{conley2008} have neglected the role of DC voltage in modelling and analysis of whirling dynamic of nano-oscillators using first-order averaging method. However, DC voltage is a prime component of electrostatic actuation and induces static pre-deflection in nano-oscillators, which breaks down the symmetry of beam displacement and becomes the source of geometric quadratic nonlinearity \cite{kozinsky2006, nayfeh1997}. Moreover, DC voltage also mistunes the planar and nonplanar natural frequencies of oscillation. Hence, in this work, we have accounted these effects of DC voltage on whirling dynamics of nano-oscillators, and have solved the nonlinear dynamics problem using a combination of numerical techniques and analytical technique second-order averaging method. Additionally, an analytical criterion for initiation of whirling motion is also provided in the form of coupled algebraic equations. We demonstrate the importance of DC voltage on whirling motion of nano-oscillators and show that DC voltage can qualitatively change the initiation pattern of whirling motion.
\section{Mathematical modelling}
Figure \ref{wfig.nanowire-model} shows the schematic diagram of a doubly-clamped nanowire oscillator of radius $ \hat R$ and length $\hat L$, and is placed on a side of an electrode plate at a gap $\hat g$. The nanowire is electrostatically actuated by applying bias voltage $\hat V$ which is a combination of DC voltage $\hat V_{DC}$ and AC voltage of amplitude $\hat V_{AC}$ at frequency $\hat \omega_f$. We use hat ($\hat .$) in the variables for differentiating them with their non-dimensional forms to be introduced later.

\indent Since the aspect ratio of nanowire oscillators is usually very high, we have employed Euler-Bernoulli beam theory for modelling of planar $\hat u(\hat x, \hat t)$ and nonplanar $\hat v(\hat x, \hat t)$ displacements corresponding  to $\hat z$ and $\hat y$ directions, respectively. The governing differential equations of motion for $\hat u(\hat x, \hat t)$ and $\hat v(\hat x, \hat t)$ are nonlinear coupled partial differential equations  \cite{ conley2008, chen2010}
\begin{equation}\label{weq.pde-equation1}
\begin{array}{l}
\displaystyle
 \hat E\hat I\frac{{\partial ^4 \hat u}}{{\partial \hat x^4 }} + \hat \rho \hat A\frac{{\partial ^2 \hat u}}{{\partial \hat t^2 }} + \hat c\frac{{\partial \hat u}}{{\partial \hat t}} = \\ \,\,\,\,\,\,\,\,\,
 \displaystyle \left[ {\frac{{\hat E\hat A}}{{2\hat L}}\int\limits_0^{\hat L} {\left( {\left( {\frac{{\partial \hat u}}{{\partial \hat x}}} \right)^2  + \left( {\frac{{\partial \hat v}}{{\partial \hat x}}} \right)^2 } \right)d\hat x + \hat N} } \right]\frac{{\partial ^2 \hat u}}{{\partial \hat x^2 }} + \hat F_e \left( {\hat u} \right), \\
 \displaystyle
 \hat E\hat I\frac{{\partial ^4 \hat v}}{{\partial \hat x^4 }} + \hat \rho \hat A\frac{{\partial ^2 \hat v}}{{\partial \hat t^2 }} + \hat c\frac{{\partial \hat v}}{{\partial \hat t}} = \\ \,\,\,\,\,\,\,\,\,
 \displaystyle \left[ {\frac{{\hat E\hat A}}{{2\hat L}}\int\limits_0^{\hat L} {\left( {\left( {\frac{{\partial \hat u}}{{\partial \hat x}}} \right)^2  + \left( {\frac{{\partial \hat v}}{{\partial \hat x}}} \right)^2 } \right)d\hat x + \hat N} } \right]\frac{{\partial ^2 \hat v}}{{\partial \hat x^2 }},\\
 \end{array}
\end{equation}
and the boundary conditions are
\begin{equation*}
\begin{array}{l}
\displaystyle
 \left. {\frac{{\partial \hat u}}{{\partial \hat x}}} \right|_{\hat x = \,0}  = \left. {\frac{{\partial \hat u}}{{\partial \hat x}}} \right|_{\hat x = \,\hat L}  = \left. {\hat u} \right|_{\hat x = \,0}  = \left. {\hat u} \right|_{\hat x = \,\hat L}  = 0,\\
 \displaystyle
 \left. {\frac{{\partial \hat v}}{{\partial \hat x}}} \right|_{\hat x = \,0}  = \left. {\frac{{\partial \hat v}}{{\partial \hat x}}} \right|_{\hat x = \,\hat L}  = \left. {\hat v} \right|_{\hat x = \,0}  = \left. {\hat v} \right|_{\hat x = \,\hat L}  = 0 \cdot
 \end{array}
\end{equation*}
In Eq. \eqref{weq.pde-equation1}, Young's modulus, mass density, cross-sectional area, and moment of inertia of the nanowire are represented by $\hat E$, $\hat \rho$, $\hat A$ and $\hat I$, respectively. It is assumed that nanowire is oscillating in a viscous medium whose damping effect on nanowire motion can be quantified with coefficient $\hat c$. The parameter $\hat N$ denotes uniform axial load to account residual stress. Equation \eqref{weq.pde-equation1} is the set of coupled partial differential equations, where coupling is due to geometric nonlinearity through beam stretching $\Delta = (1/2)\int_0^{\hat L} \left(  \hat u^{{'}^2} +  \hat v^{{'}^2}  \right) d\hat x$. Nonlinear electrostatic driving force $\hat F_e (\hat u)$ is generated by applying bias voltage $\hat V$ between the plate electrode and the nanowire as \cite{ouakad2010, bhushan2011}
\begin{equation*}
\hat F_e (\hat u) = \frac{{\pi \varepsilon _0 \varepsilon _r \left( {\hat V_{DC}  + \hat V_{AC} \cos (\hat \omega _f \hat t)} \right)^2 }}{{\sqrt {\left( {\hat g + \hat R - \hat u} \right)^2  - \hat R^2 } \left[ {\cosh ^{ - 1} \left( {\frac{{\hat g + \hat R - \hat u}}{{\hat R}}} \right)} \right]^2 }} \cdot
\end{equation*}
Here, $\varepsilon_0$ and $\varepsilon_r$ are vacuum and relative permittivities; the value of  $\varepsilon_r$ has been taken as unity. \\
\indent Equation \ref{weq.pde-equation1} has been transformed into non-dimensional form for simplified analysis. The non-dimensional form of planar displacement $u(x, t)$, non-planar displacement  $v(x, t)$, spatial coordinate $x$, and time $t$ are related to their dimensional form as
\begin{equation}\label{weq.nondimensional-variable}
u = \frac{{\hat u}}{{\hat g}},\,\,v = \frac{{\hat v}}{{\hat g}},\,\,x = \frac{{\hat x}}{{\hat L}},\,\,\text{and} \,\,t = \frac{{\hat t}}{{\hat T}} \cdot
\end{equation}
Here, $\hat T$ is a time constant whose expression is given in Eq. \eqref{weq.pde-parameters}. The non-dimensional form of Eq. \eqref{weq.pde-equation1} has been obtained through substituting (\ref{weq.nondimensional-variable}) in Eq. \eqref{weq.pde-equation1} and is given as
\begin{equation} \label{weq.pde-equation2}
\begin{array}{l}
 u^{''''}  + \ddot u + c\dot u = \left[ {\alpha _1 \int\limits_0^1 {\left( {u^{'^2 }  + v^{'^2 } } \right)dx} } + N\right]u^{''}  + \alpha _2 V^2 f_e(u), \\
 v^{''''}  + \ddot v + c\dot v = \left[ {\alpha _1 \int\limits_0^1 {\left( {u^{'^2 }  + v^{'^2 } } \right)dx} } + N\right]v^{''}, \\
 \end{array}
\end{equation}
where the boundary conditions are
\begin{equation*}
\begin{array}{l}
 \left. {u'} \right|_{x = \,0}  = \left. {u'} \right|_{x = 1}  = \left. u \right|_{x = \,0}  = \left. u \right|_{x = \,1}  = 0,\\
 \left. {v'} \right|_{x = \,0}  = \left. {v'} \right|_{x = 1}  = \left. v \right|_{x = \,0}  = \left. v \right|_{x = \,1}  = 0 \cdot
 \end{array}
\end{equation*}
Definitions of various terms of Eq. \eqref{weq.pde-equation2} are
\begin{equation}\label{weq.pde-parameters}
\begin{array}{l}
\displaystyle
 f_e (u) = \frac{\pi }{{\sqrt {\left( {1 + R_0  - u} \right)^2  - R_0 ^2 } \left[ {\cosh ^{ - 1} \left( {\frac{{1 + R_0  - u}}{{R_0 }}} \right)} \right]^2 }},\, \\
\displaystyle c = \frac{{\hat cL^4 }}{{\hat E\hat I\hat T}},\,\,\,\alpha _1  = \frac{{\hat A\hat g^2 }}{{2\hat I}},\,\,\,\alpha _2  = \frac{{\varepsilon _0 \varepsilon _r \hat L^4}}{{\hat g^2 \hat E\hat I}},\,\, \\ \displaystyle
N = \frac{{\hat N\hat L^2 }}{{\hat E\hat I}},\,R_0  = \frac{{\hat R}}{{\hat g}},\,\,\,\hat T = \sqrt {\frac{{\hat \rho \hat A\hat L^4 }}{{\hat E\hat I}}} ,\,\,\, \\ \displaystyle
 V = \left( {V_{DC} + V_{AC} \cos (\omega _f t)} \right),\,\,\,{\rm{and}}\,\,\omega _f  = \hat \omega _f \hat T \cdot \\
\end{array}
\end{equation}
In Eq. \eqref{weq.pde-equation2}, we use $(^{'})$ and $(^{.})$ to denote partial derivatives with respect to $x$ and $t$ respectively and retain $V_{DC}$ and $V_{AC}$ in dimensional-form even after removing hat ($\hat .$). Next section presents dynamical equations of motion for whirling dynamics which have been derived using Galerkin based reduced order modelling technique.
\section{Reduced order modelling}
\begin{figure*}
\centering
\includegraphics[width=7.5cm]{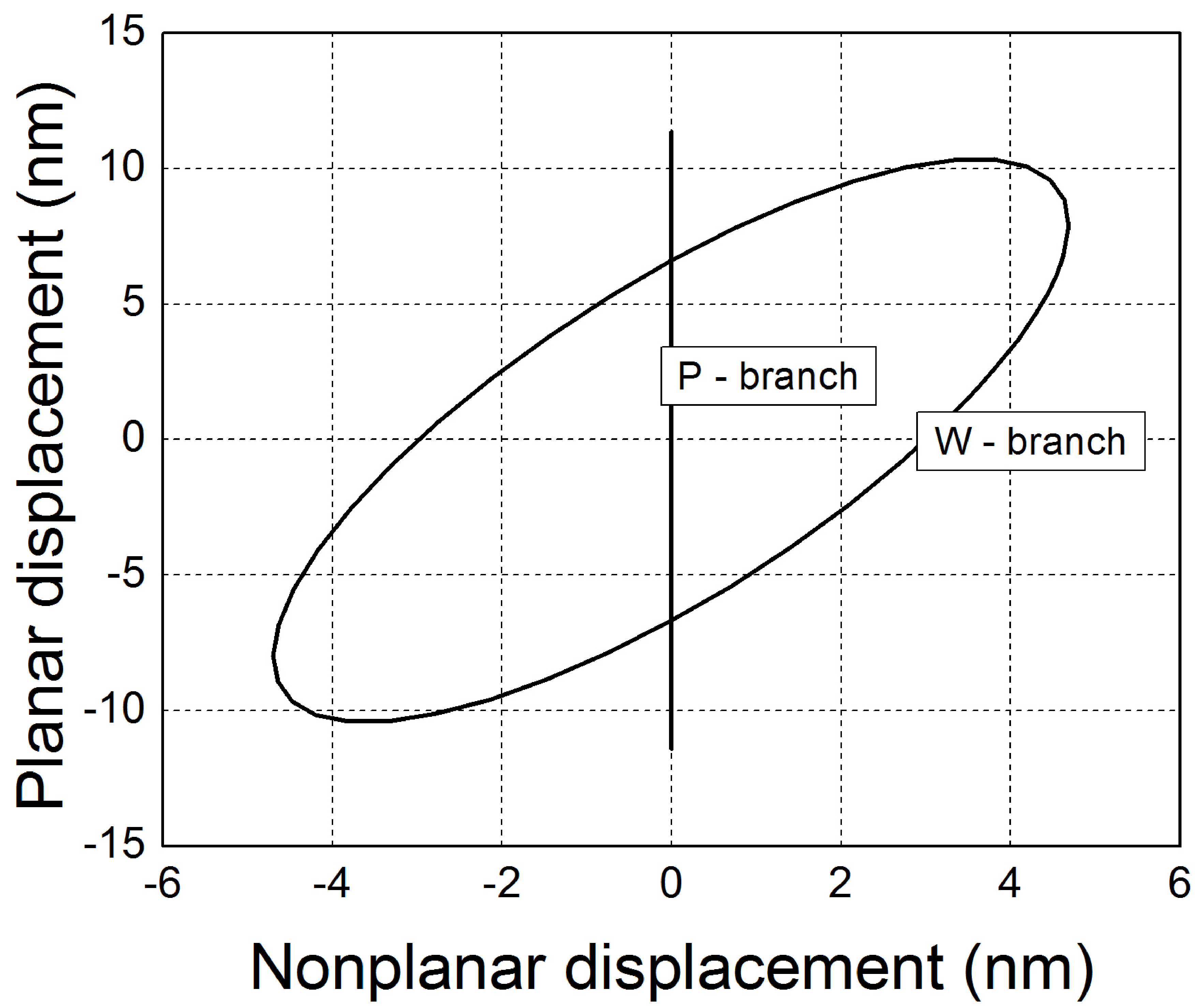}
\caption{Trajectories of mid-point of the nanowire oscillator during planar and whirling motion.}
\label{wfig.trajectory}
\end{figure*}

\begin{figure*}
\captionsetup[subfigure]{labelformat=empty}
\subfloat[]{\includegraphics[width=7.5cm]{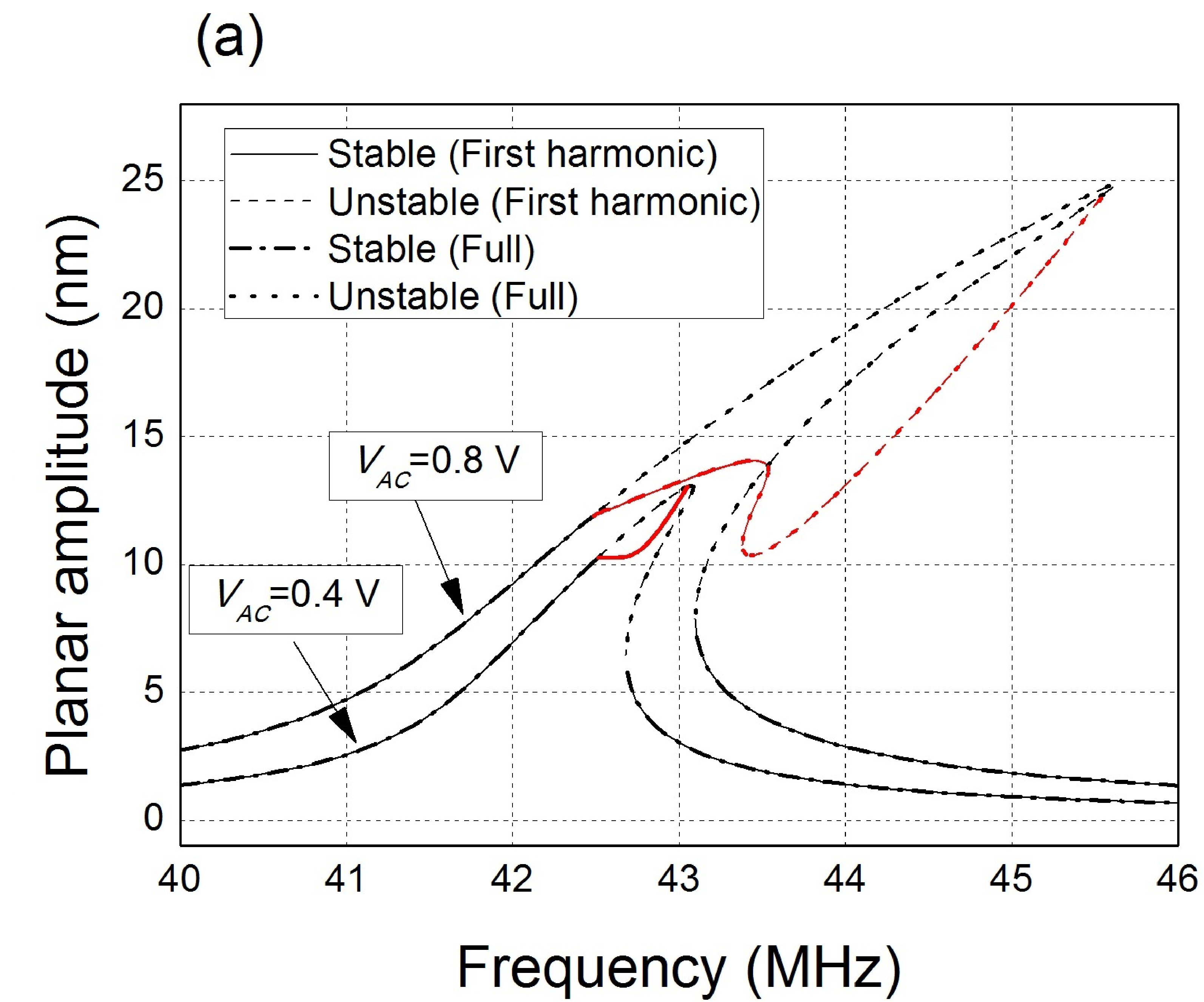}}
\hspace{0.5cm} 
\subfloat[]{\includegraphics[width=7.5cm]{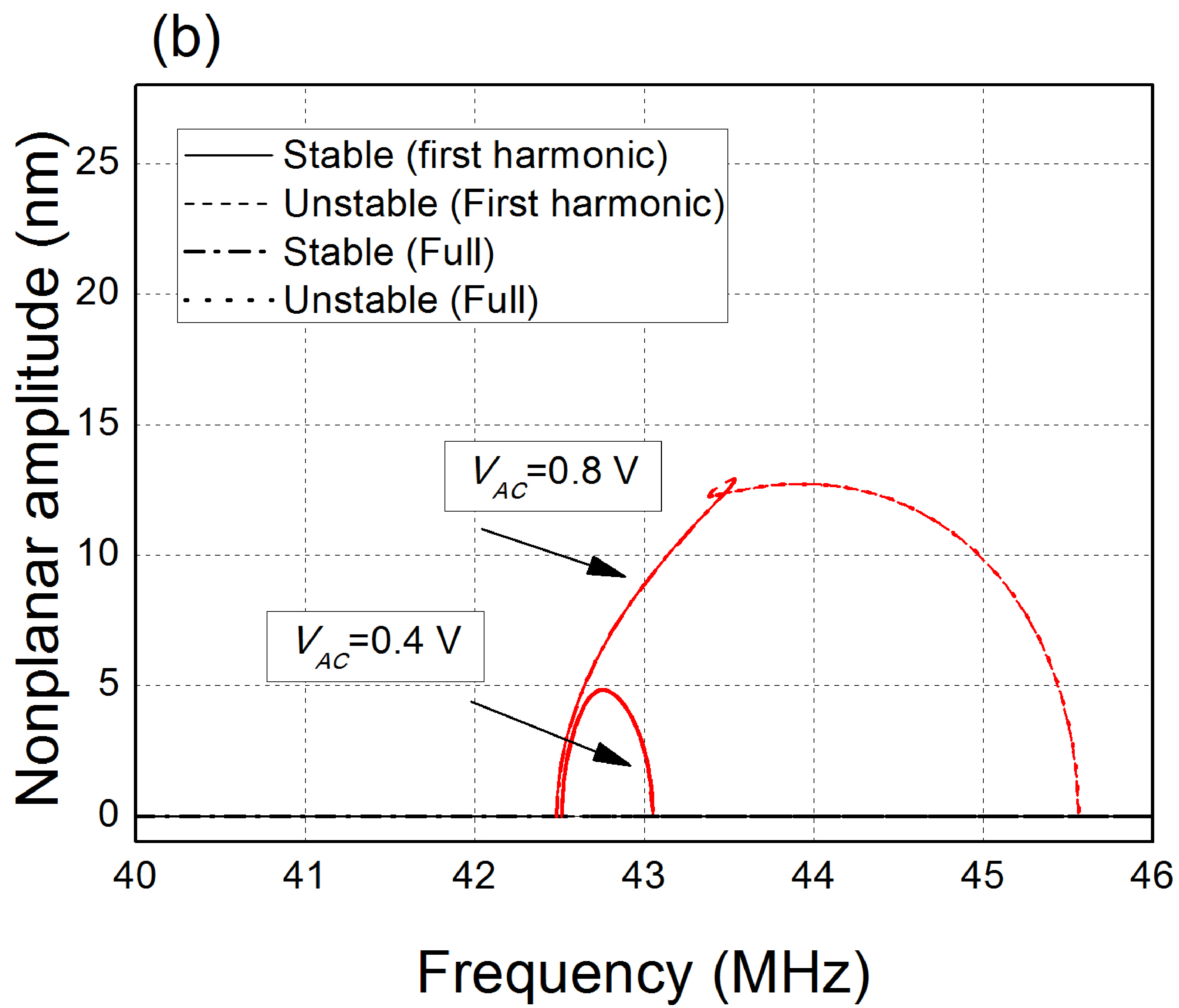}}
\caption{(a) Planar resonance curves and (b) nonplanar resonance curves obtained by solving single-mode ROM for $V_{DC}=5$ V and two different values of $V_{AC}$ to show relative importance of first direct harmonic term in full dynamic excitation.}
\label{wfig.para}
\end{figure*}

To study whirling dynamics of a nanowire oscillator, we have adopted Galerkin based reduced order modelling technique for reducing governing partial differential equations of motion \eqref{weq.pde-equation2} into system of ordinary differential equations \cite{rasekh2010, bhushan2011, younis20031}. To do so, planar and nonplanar displacements have been assumed as
 \begin{equation}\label{weq.assumed-sol}
u(x,t) = \sum\limits_{i = 1}^p {\phi _i (x)u_i (t)} \,\,\,\,{\rm{and}}\,\,\,\,v(x,t) = \sum\limits_{i = 1}^p {\phi _i (x)v_i (t)} \cdot
 \end{equation}
Here, $\phi_i(x)$ is normalised mode shape ($\int_0^1 \phi_i^2 dx=1$) of a doubly-clamped straight beam, whereas $u_i(t)$ and $v_i(t)$ are temporal modal coordinates. We can calculate the mode shape $\phi_i(x)$ and the corresponding natural frequency $\omega_i$ by solving equation \cite{rao2004}
\begin{equation}\label{peq.ode-modal}
\phi ^{''''}  = N\phi ^{''}  + \omega ^2 \phi ,
\end{equation}
where the boundary conditions are
\begin{equation*}
\phi  |_{x \,= \,0}  = \phi |_{x\, = \,1}  = \phi^{'}|_{x\, = \,0}  = \phi^{'} |_{x \,= \,1}  = 0\cdot
\end{equation*}
After substituting the assumed form of solutions (\ref{weq.assumed-sol})  for $u$ and $v$ in Eq. \eqref{weq.pde-equation2}, multiplying the outcome with $\phi_i(x)$, and integrating the equations from $x=0$ to $x=1$, we have obtained the reduced order model (ROM) as\\
\begin{equation} \label{weq.rom-multi}
\begin{array}{l}
 \ddot u_n  + c\dot u_n  + \omega _n^2 u_n  = \\ \alpha _1
 \left( {\sum\limits_{i,\,j,\,k\, = 1}^p {\left( {u_i u_j u_k  + v_i v_j u_k } \right)\int\limits_0^1 {\left( {\phi _i^{'} \phi _j^{'} } \right)} dx\,\cdot\int\limits_0^1 {\left( {\phi _k^{''} \phi _n } \right)} dx} } \right)\,
 \\+ \alpha _2 V^2 \int\limits_0^1 {f_e \left( {\sum\limits_{i = 1}^p {\phi _i u_i } } \right)\phi _n } dx, \\
 \ddot v_n  + c\dot v_n  + \omega _n^2 v_n  = \\ \alpha _1
 \left( { \sum\limits_{i,\,j,\,k\, = 1}^p {\left( {u_i u_j v_k  + v_i v_j v_k } \right)\int\limits_0^1 {\left( {\phi _i^{'} \phi _j^{'} } \right)} dx\,\cdot\int\limits_0^1 {\left( {\phi _k^{''} \phi _n } \right)} dx} } \right), \\
 {\text{for } n = 1,\,\, 2,\,\, 3....p} \cdot \\
 \end{array}
\end{equation}
The resulting ROM is a system of ordinary differential equation of initial value problem and can be considered as dynamical equations of motion of a multi-degree freedom system.\\
\indent In this work, we have numerically and analytically investigated a typical silicon nanowire oscillator of length $\hat L$ = 3000 nm and radius $\hat R$ = 25 nm. The properties of the investigated nanowire oscillator have been chosen nearby the properties of the experimentally investigated nanowire oscillators of other reasearchers \cite{solanki2010, zhu2009}. The gap $\hat g$ between nanowire and electrode plate has been chosen as 300 nm and quality factor $Q$ = 100 has been taken to account for damping effects. As in a linear harmonic oscillator, quality factor $Q$ is related to the non-dimensional damping coefficient $c$ as $Q=\omega_1 / c$. Further, unless otherwise specified magnitude of axial load $\hat N$ is zero. \\
\indent In order to infer the interaction between various planar and nonplanar modes of vibration of the nanowire through internal resonance, we first study the variation of planar and nonplanar natural frequencies with respect to DC voltage  \cite{nayfeh1979, gutschmidt}. To do so, the ROM (\ref{weq.rom-multi}) has been re-written as a system of second order ordinary differential equations for free undamped vibration after neglecting damping and AC voltage terms as $\mathbf{\ddot  U} = \mathbf{F}(\mathbf{U})$, where $\mathbf{F}$ is a vector-valued function of variable $\mathbf{U} = {\left[ {{u_1}\,{u_2}...{u_p}...{v_1}\,{v_2}...{v_p}}\right ]^T}$. After applying DC voltage, the nanowire acquires static equilibrium position $\mathbf{U_s}$ which is the solution of system of nonlinear algebraic equations $\mathbf{F(U_s) = 0}$. The linearised dynamical equations at static equilibrium position $\mathbf{U_s}$ can be expressed as $\mathbf{\ddot W(t)} = \mathbf{J}(\mathbf{U_s})\mathbf{W}(t),$ where  $\mathbf{W}(t)$ is a translated coordinate such that $\mathbf{W}(t) = \mathbf{U}(t)-\mathbf{U_s}$ and $\mathbf{J}$ is the Jacobian matrix of $\mathbf{F}$. The square root of eigenvalues of Jacobian matrix $\mathbf{J}$ are the natural frequencies; and the mode shapes can be calculated from eigenvectors of $\mathbf{J}$.\\
\begin{figure*}
\captionsetup[subfigure]{labelformat=empty}
\subfloat[]{\includegraphics[width=7.5cm]{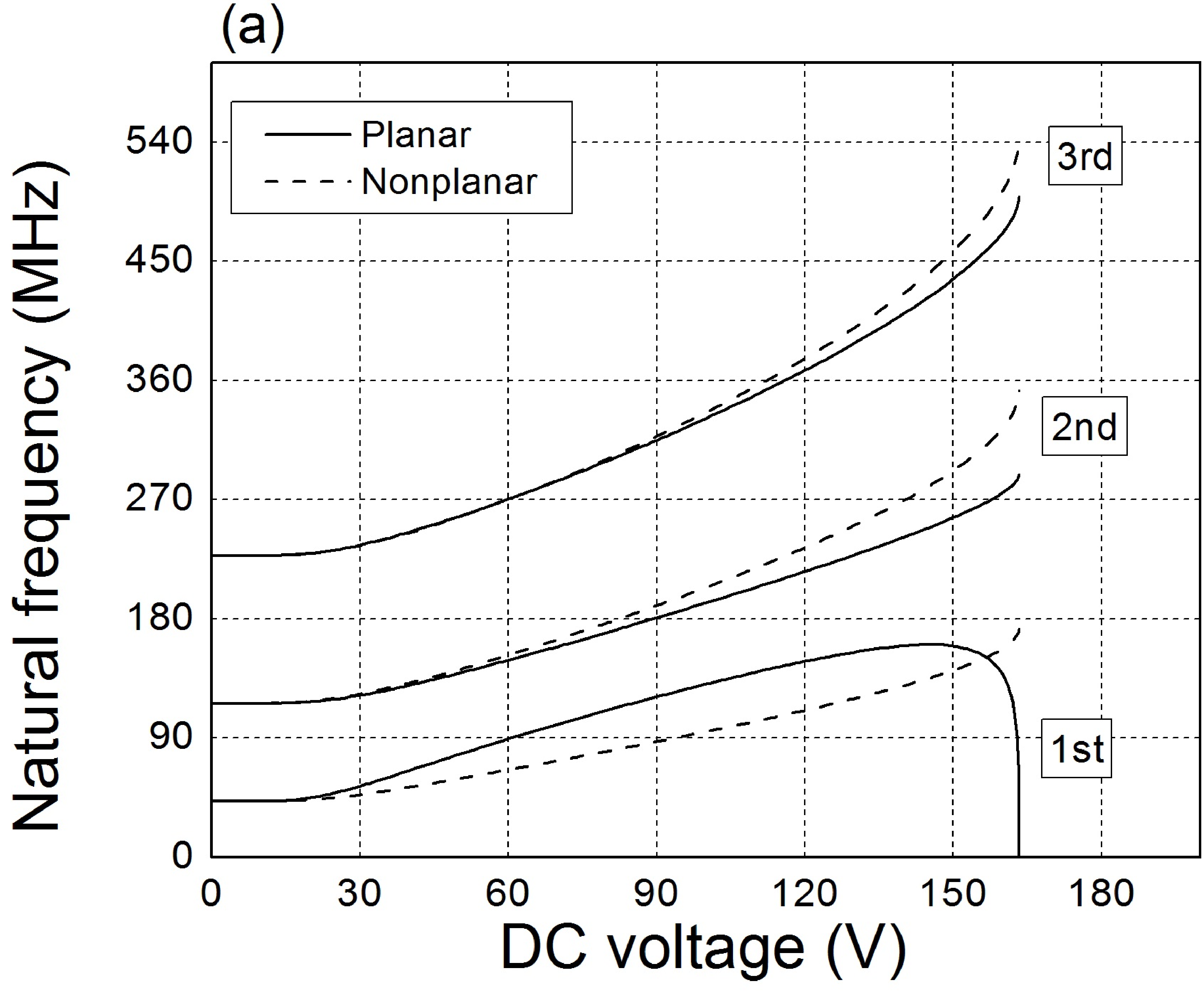}}
\hspace{0.5cm} 
\subfloat[]{\includegraphics[width=7.5cm]{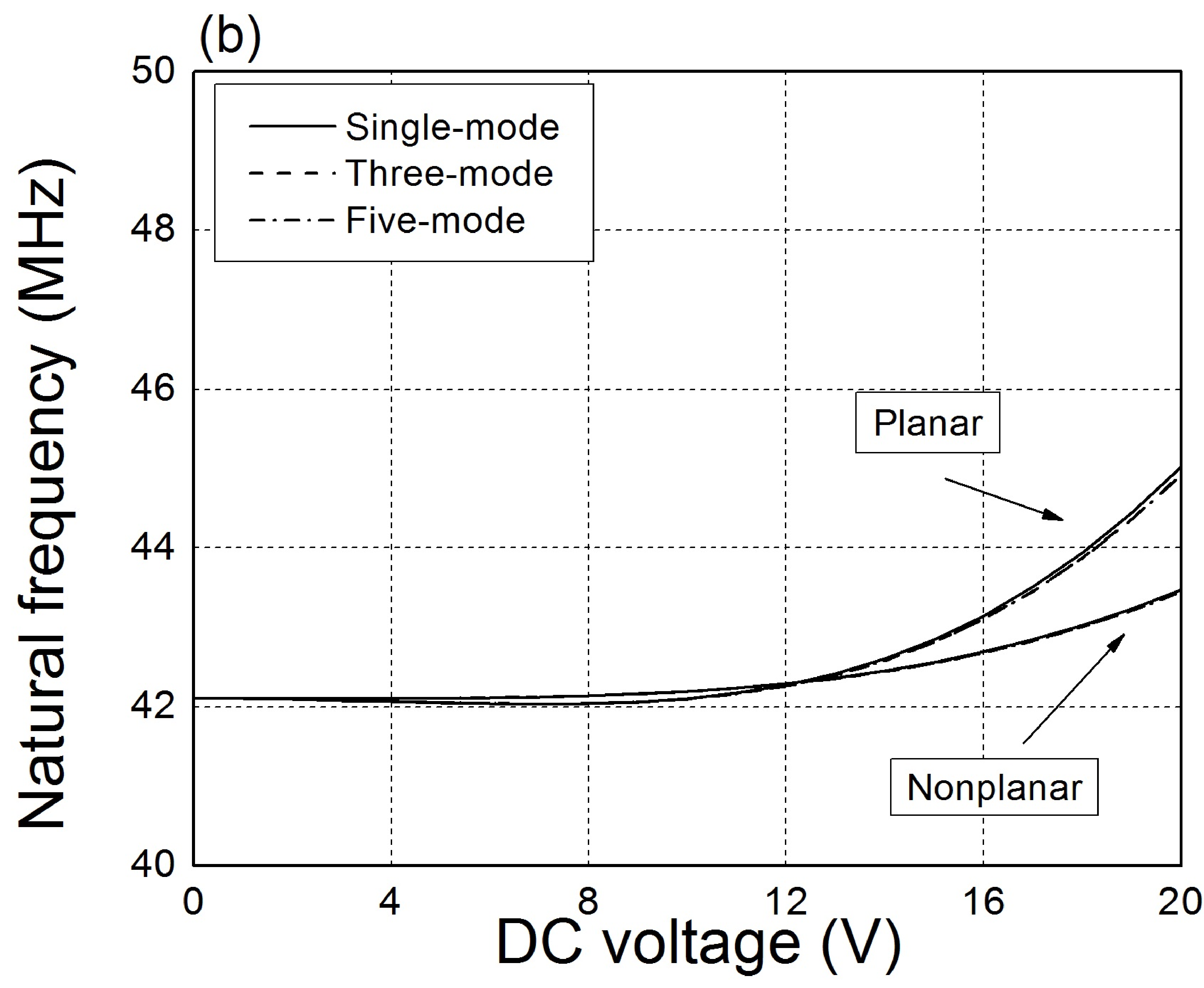}}
\caption{Effects of DC voltage on (a) variation of first three planar and nonplanar natural frequencies till static pull-in voltage and (b) variation of first planar and nonplanar natural frequencies till $V_{DC}$ = 20 V.}
\label{wfig.stat-eigen}
\end{figure*}
\indent Figure \ref{wfig.stat-eigen}(a) shows the variation of first three planar and nonplanar natural frequencies with respect to change in magnitude of DC voltage. These results have been obtained by numerically solving $\mathbf{F(U_s) = 0}$ and calculating eigenvalues of $\mathbf{J}$ corresponding to five-mode ROM (Eq. \eqref{weq.rom-multi} with $p$ = 5)  in MATLAB environment. From the figure, we can observe that first planar natural frequency initially shows little variation till $V_{DC}= 20$ V. As the magnitude of $V_{DC}$ is increased further, this frequency first increases and then decreases to zero at static pull-in voltage. We can further observe that natural frequencies of second and third planar as well as all nonplanar modes of vibration show monotonous rise with increment in magnitude of $V_{DC}$.\\
\indent  In the present investigation, in addition to numerical study, we have analytically investigated resonance behaviour of the nanowire oscillator using averaging method by considering the system as weakly nonlinear. We have considered the resonance of the nanowire oscillator around first natural frequency and limited our study till $V_{DC}=20$ V, where first, second, and third natural frequencies (both planar and nonplanar) are around 42, 116, and 227 MHz respectively. Since, internal resonance in a dynamical system may occur when some frequencies are commensurable or nearly commensurable \cite{nayfeh1979}, the possibility of internal resonance in our system, within the scope of this investigation, is between first planar and first nonplanar modes of vibration through one-to-one internal resonance. We demonstrate further in this paper that planar to whirling motion transition occurs because of interaction of these two modes of vibration. Figure \ref{wfig.stat-eigen}(b) shows the variation of first planar and nonplanar natural frequencies till $V_{DC}$ = 20 V. We have observed good agreement between numerical solutions of single-mode ROM (Eq. \eqref{weq.rom-multi} with $p$ = 1) and ROMs having higher modes. We can also observe in this figure that there is minor shift of planar and nonplanar natural frequencies with variation of DC voltage in range $V_{DC} = 0$ V to $V_{DC} = 20$ V, and also in this range of DC voltage, the static deflection of the nanowire is very small (mid-point deflection of the nanowire is less than 0.5\% of the length of the nanowire).\\
\indent The single-mode ROM is a two degree of freedom system, and has been deduced from Eq. \eqref{weq.rom-multi} to study interaction of first planar and nonplanar modes of vibration as
\begin{equation} \label{weq.rom1}
\begin{array}{l}
 \ddot u_1  + c\dot u_1  + \omega _1^2 u_1  + \alpha _1 \left( {\int\limits_0^1 {\phi _1^{'^2 } dx} \,} \right)^2 u_1 \left( {u_1^2  + v_1^2 } \right) = \\ \alpha _2 V^2 \int\limits_0^1 {f_e \left( {\phi _1 u_1 } \right)\phi _1 } dx, \\
 \ddot v_1  + c\dot v_1  + \omega _1^2 v_1  + \alpha _1 \left( {\int\limits_0^1 {\phi _1^{'^2 } dx} \,} \right)^2 v_1 \left( {u_1^2  + v_1^2 } \right) = 0 \cdot\\
 \end{array}
\end{equation}
The single-mode ROM has been further modified by translating planar $u_1(t)$ and nonplanar $v_1(t)$ displacement components to static equilibrium position. Planar component $u_s$ of static equilibrium position can be calculated by solving Eq. \eqref{weq.rom1} after setting the AC voltage and time derivative terms to zero. Since, actuation voltages are present only in the planar direction; static displacement component in nonplanar direction is zero. The origin of displacement co-ordinates $[u_1(t)\,\, v_1(t)]$ of Eq. \eqref{weq.rom1} is then shifted to static equilibrium position $[u_s\,\, 0]$. Now planar and nonplanar displacement co-ordinates at static equilibrium point become $u_d(t) = u_1(t) - u_s$ and $v_d(t) = v_1(t)$, and the modified form of single-mode ROM transforms to
\begin{equation}\label{weq.rom2}
\begin{array}{l}
 \ddot u_d  + c\dot u_d  + k_{1u} u_d  + k_{2u} u_d^2  + k_{2v} v_d^2  + k_{3u} u_d^3  + k_{3uv} u_d v_d^2\\
 = \left( {V_{AC}^2 \cos ^2 (\omega _f t) + 2 V_{DC} V_{AC} \cos (\omega _f t)} \right) {\sum\limits_{i = 0}^3 {C_i u_d^i } }, \\
 \ddot v_d  + c\dot v_d  + k_{1v} v_d  + k_{2uv} u_d v_d  + k_{3v} v_d^3  + k_{3uv} u_d^2 v_d  = 0 \cdot\\
 \end{array}
\end{equation}
 Various parameters of Eq. \eqref{weq.rom2} are defined as
\begin{equation}\label{weq.nonlinear-parameter1}
\begin{array}{l}
 k_{1u}  = \omega _1^2  + 3k_3 u_s^2  - C_1 ,\,\,k_{1v}  = \omega _1^2  + k_3 u_s^2 ,\,\,
 k_{2u}  = 3k_2 - C_2, \\ k_{2v}  = k_2 ,\,\,k_{2uv}  = 2 k_2, \,\,
 k_{3u}  = k_3  - C_3 ,\,\,k_{3v}  = k_{3uv}  = k_3, \\ \text{where} \,\,\, k_2=k_3 u_s,\,\,\,k_3  = \alpha _1 \left( {\int\limits_0^1 {\phi _1^{'^2 } } dx\,} \right)^2 ,\,\,
 \\
 \text{and}\,\,C_i  = \frac{1}{{i!}}\alpha _2 \int\limits_0^1 {f_e^{(i)} (u_s \phi _1 )\phi _1^{i + 1} dx} ,\,\,\,\,i = 0,\,1,\,2,\,3 \cdot  \\
 \end{array}
\end{equation}
In Eq. \eqref{weq.rom2}, we have expressed electrostatic forcing function $f_e(u)$ of (\ref{weq.pde-parameters}) in Taylor series of third order. Note that, Eq. \eqref{weq.rom2} contains both symmetric odd order and asymmetric even order nonlinearities. The parameter $C_i$ arises due to nonlinear electrostatic actuation force, whereas $k_2$ and $k_3$ arise due to beam stretching. An important point to note here is that $k_2=k_3 u_s$ results from the static deflection by applied DC voltage and is the source of geometric quadratic nonlinearities in the coupled oscillator.\\
\begin{figure*}
\captionsetup[subfigure]{labelformat=empty}
\subfloat[]{\includegraphics[width=7.5cm]{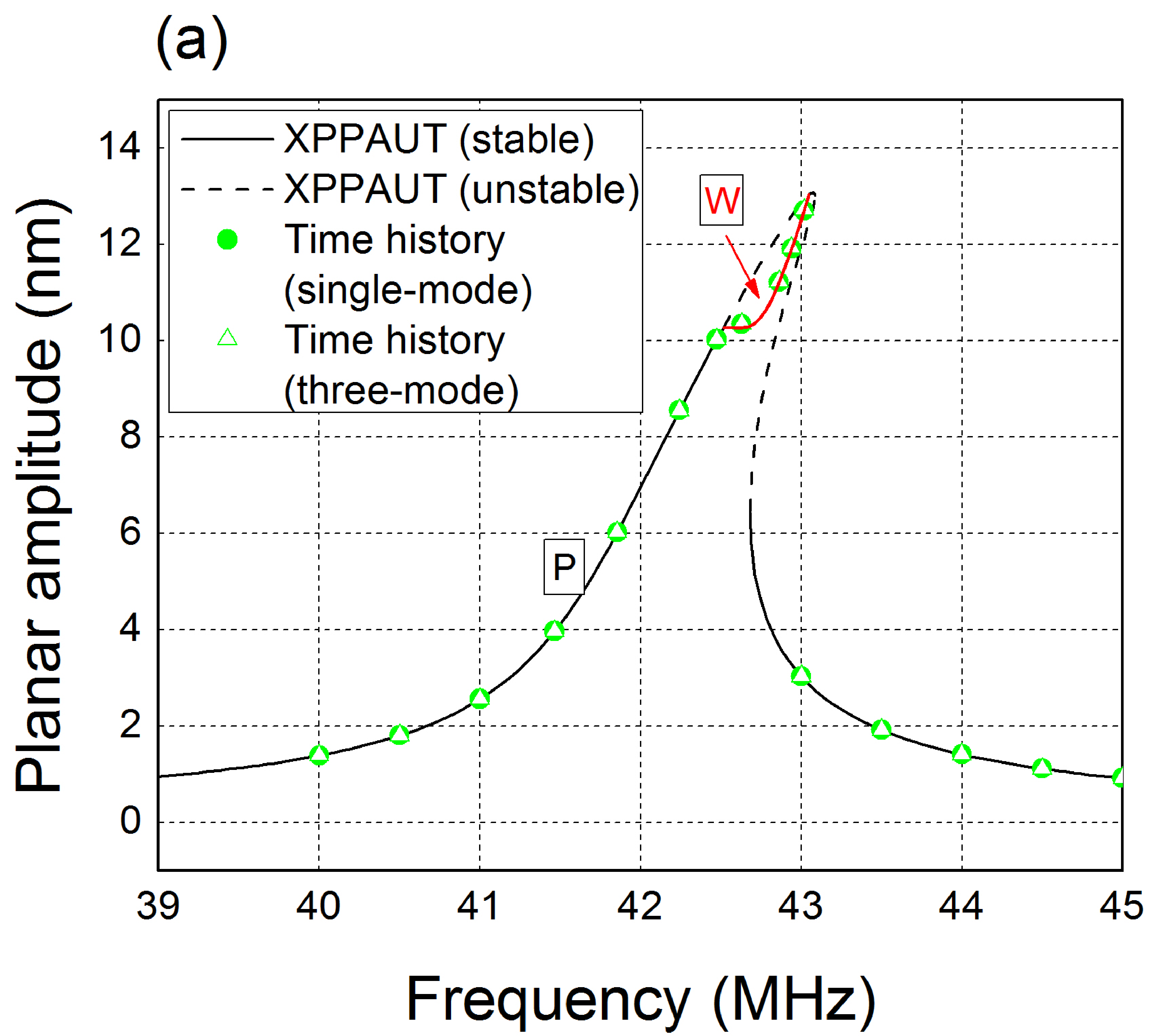}}
\hspace{0.5cm} 
\subfloat[]{\includegraphics[width=7.5cm]{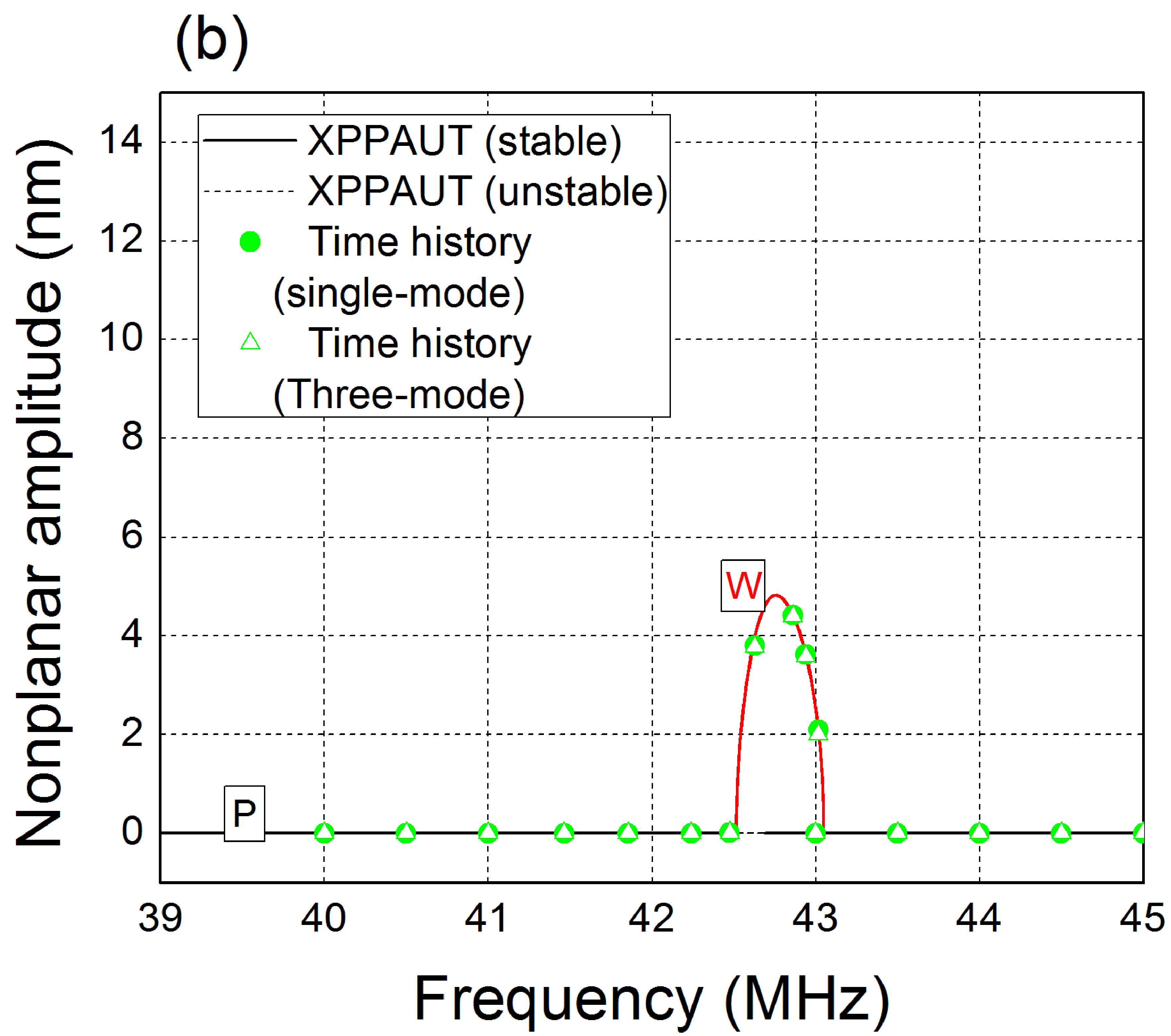}}
\caption{Comparison of (a) planar resonance curves and (b) nonplanar resonance curves obtained using single-mode and multi-mode ROM for $V_{DC}=5$ V and $V_{AC}=0.40$ V.}
\label{wfig.rc-5-0p4}
\end{figure*}
\indent Single-mode ROM \eqref{weq.rom2} is the basic dynamical equation of motion which has been solved using numerical and analytical techniques. First, the numerical solutions of single-mode ROM \eqref{weq.rom2} have been compared with solutions of three-mode ROM (Eq. \eqref{weq.rom-multi} with $m$ = 3) to validate the solutions of single-mode ROM. The comparison is shown here in Figs. \ref{wfig.rc-5-0p4}(a) and \ref{wfig.rc-5-0p4}(b) depicting resonance curves for oscillation near first natural frequency for $V_{DC} = 5$ V and $V_{AC}$ = 0.40 V. Specifically, Figs. \ref{wfig.rc-5-0p4}(a) and \ref{wfig.rc-5-0p4}(b) show variation of planar and nonplanar mid-point amplitudes of the nanowire oscillator, respectively, with forcing frequency $ \omega_f$, and are termed here as planar and nonplanar resonance curves. In these figures, we compare solutions of single-mode ROM (circle) and three-mode ROM (triangle) corresponding to Eq. \eqref{weq.rom-multi} with the solutions of single-mode ROM \eqref{weq.rom2} (line). The single- and three-mode ROMs corresponding to Eq. \eqref{weq.rom-multi} have been solved in MATLAB environment by integrating the equations for long-time to get periodic solutions. We have solved the equations for both forward and reverse sweeps with respect to small increment of $\omega_f$ to obtain resonance curves. It may be noted here that, for solving the equations of a particular value of $\omega_f$, the initial condition has been provided corresponding to the periodic solution of last simulated $\omega_f$ with small perturbation in nonplanar coordinate $v_i(t)$. This particular step is essential to achieve planar to whirling motion transition in numerical simulations, and is consistent with experiments where small perturbations are always present because of interaction with environment. Single-mode ROM \eqref{weq.rom2} has been numerically solved using nonlinear dynamic software XPPAUT \cite{ermentrout2002} to obtain periodic solutions and their stability in continuation of $\omega_f$; continuous lines in Figs. \ref{wfig.rc-5-0p4}(a) and \ref{wfig.rc-5-0p4}(b) represent stable solutions, whereas dashed lines represent unstable solutions. As can be observed from the figures, solutions of Eqs. \eqref{weq.rom2} and \eqref{weq.rom-multi} are in good agreement and it demonstrates that single-mode ROM (\ref{weq.rom2}) is capable to capture whirling dynamic behaviour of the nanowire oscillator. In Figs. \ref{wfig.rc-5-0p4}(a) and \ref{wfig.rc-5-0p4}(b), two distinct branches of both planar and nonplanar resonance curves are labelled with letters P and W which denote P-branch and W-branch, respectively. When nanowire oscillation is characterised by P-branch, the nanowire oscillates only in planar direction, i.e., absence of nonplanar motion ($v_d(t) = 0$). The W-branch indicates oscillation with finite magnitude of both planar and nonplanar amplitudes; whirling motion can exists on this branch only. It is interesting, the reason behind the initiation of whirling motion can be explained using the theory of Mathieu oscillator \cite{harrison1948}. Typical trajectories of mid-point of the nanowire in Y-Z plane (refer Fig. \ref{wfig.nanowire-model}) during planar and whirling motion are shown in Fig. \ref{wfig.trajectory}. As expected, trajectory of planar motion is straight line and whirling motion is elliptical.\\
\indent It is needful to mention here that we have retained only first direct harmonic excitation term $2V_{DC}V_{AC}C_0 \cos(\omega_f t)$ in Eq. \eqref{weq.rom2} and ignored other parametric excitation and second harmonic terms during solving single-mode ROM in this paper. This is because, only first direct harmonic excitation is dominant in primary resonance near first natural frequency \cite{bhushan2014}. We show this observation here in Figs. \ref{wfig.para}(a) and \ref{wfig.para}(b) by comparing two different sets of planar and nonplanar resonance curves. One set is the solution of Eq. \eqref{weq.rom2} with full dynamic excitation, and other set is solution of Eq. \eqref{weq.rom2} with only first direct harmonic excitation $2V_{DC}V_{AC}C_0 \cos(\omega_f t)$. In Figs. \ref{wfig.para}(a) and \ref{wfig.para}(b), curves are plotted for $V_{DC} = 5$ V and two different values of AC voltage $V_{AC} = 0.4$ V and $V_{AC} = 0.8$ V; we can observe a very good agreement between the two sets of resonance curves. Hence, we can infer that the first direct harmonic excitation is dominant part of dynamic excitation, and the effect of other second harmonic and parametric terms are very small when DC voltage is much greater than AC voltage.
\section{Effects of DC voltage on resonance curves}\label{ws.dc-voltage}
\begin{figure*}
\captionsetup[subfigure]{labelformat=empty}
\subfloat[]{\includegraphics[width=7.5cm]{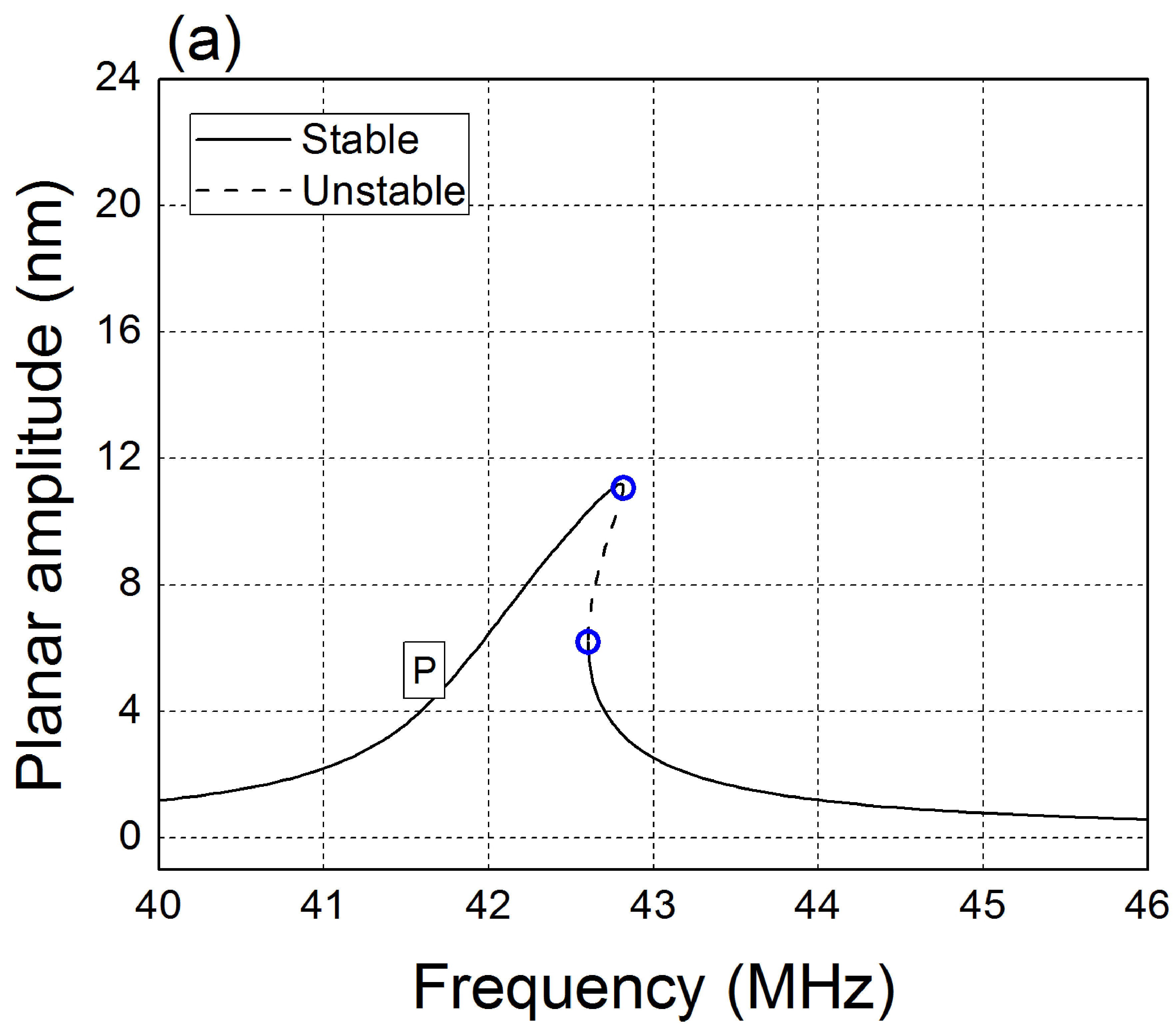}}
\hspace{0.5cm} 
\subfloat[]{\includegraphics[width=7.5cm]{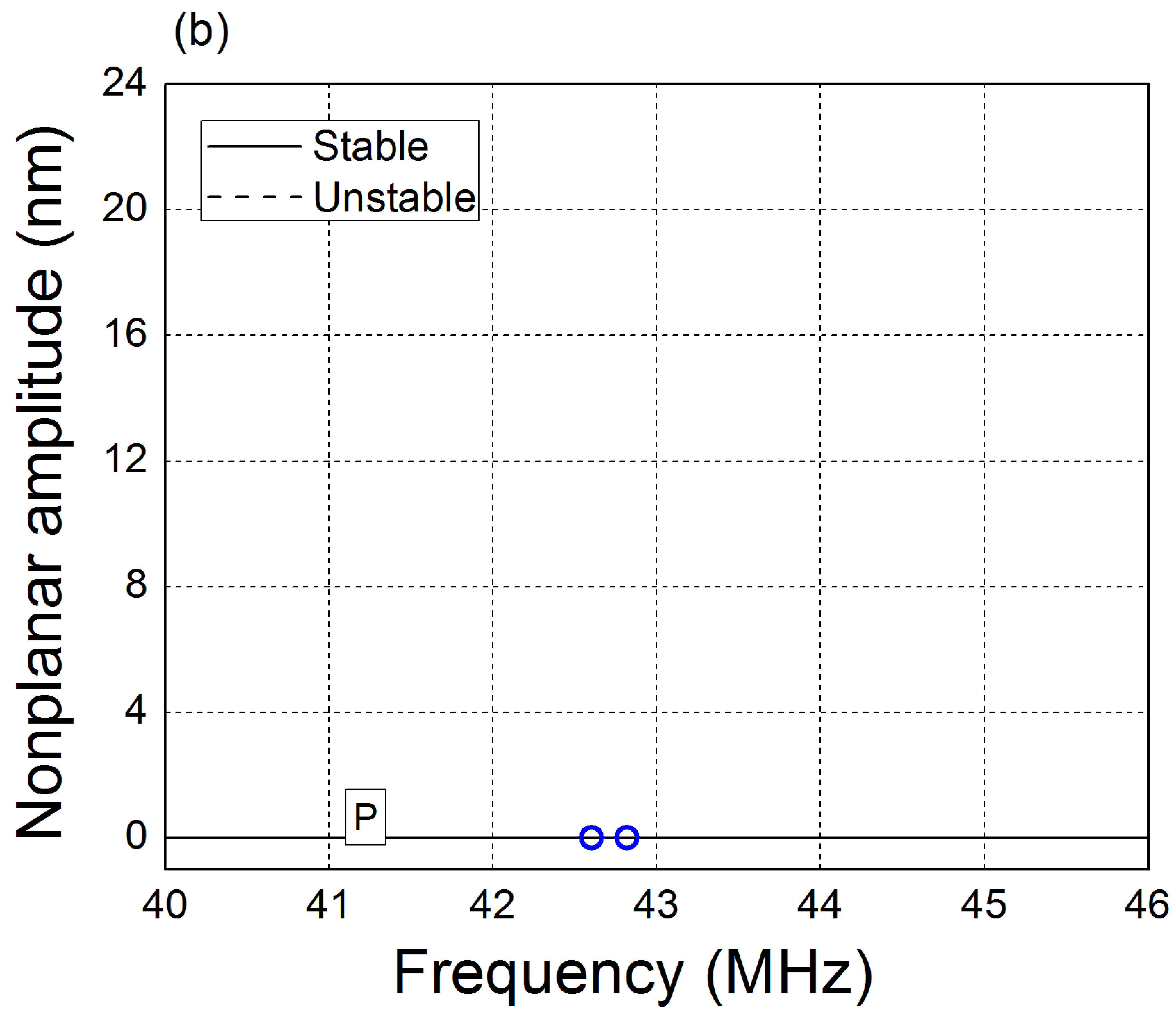}}\\[-1cm]
\subfloat[]{\includegraphics[width=7.5cm]{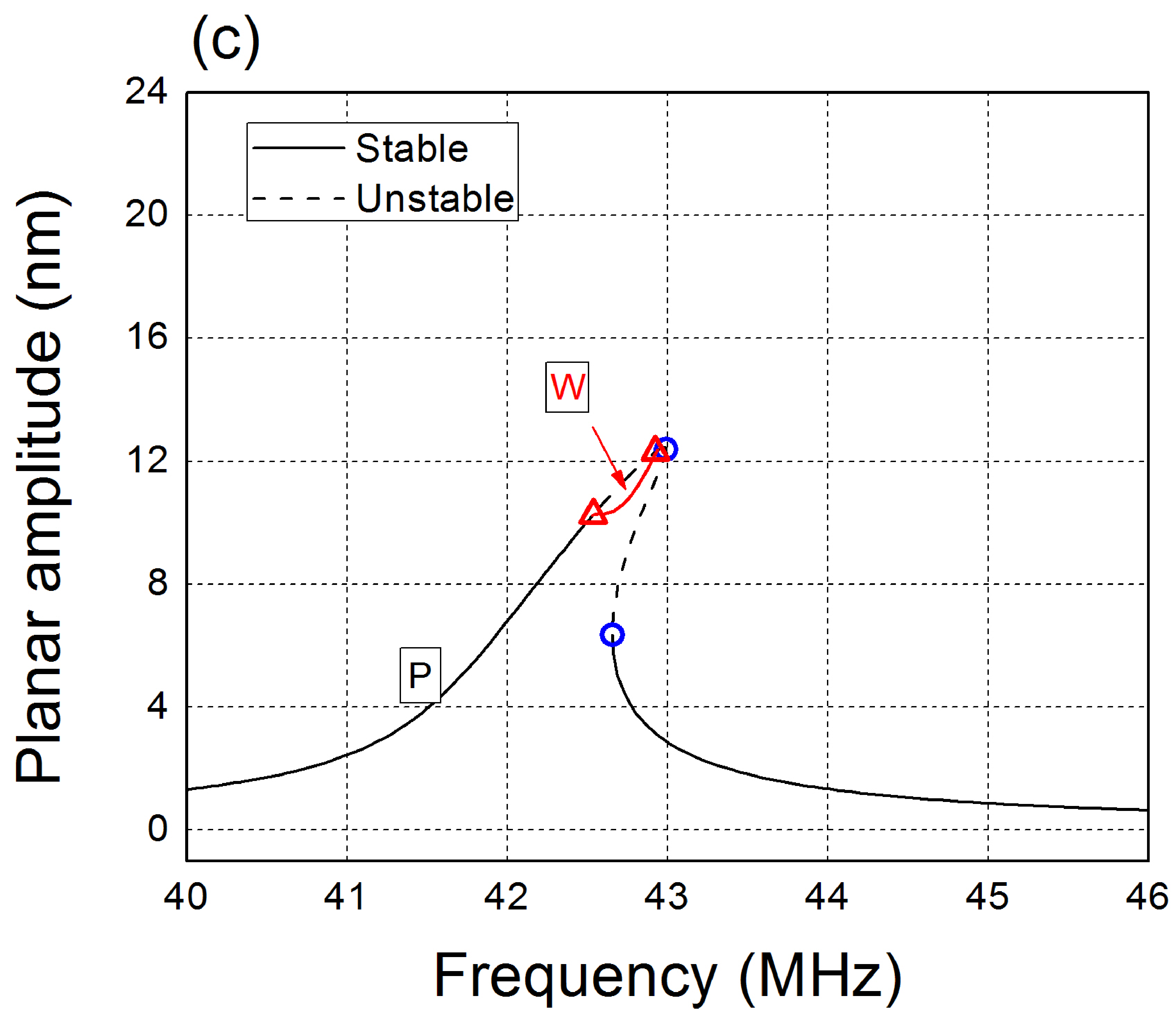}}
\hspace{0.5cm} 
\subfloat[]{\includegraphics[width=7.5cm]{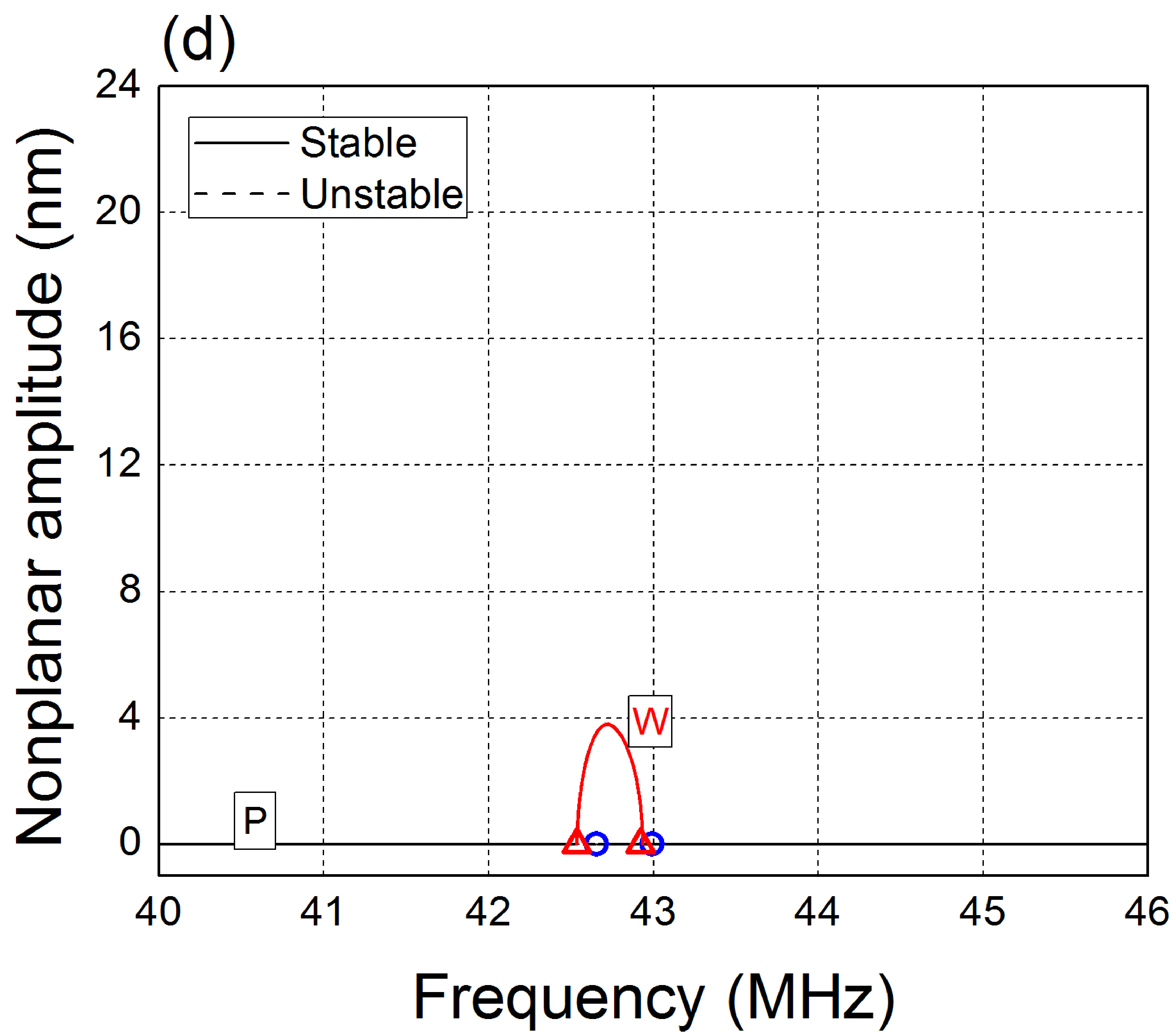}}\\[-1cm]
\subfloat[]{\includegraphics[width=7.5cm]{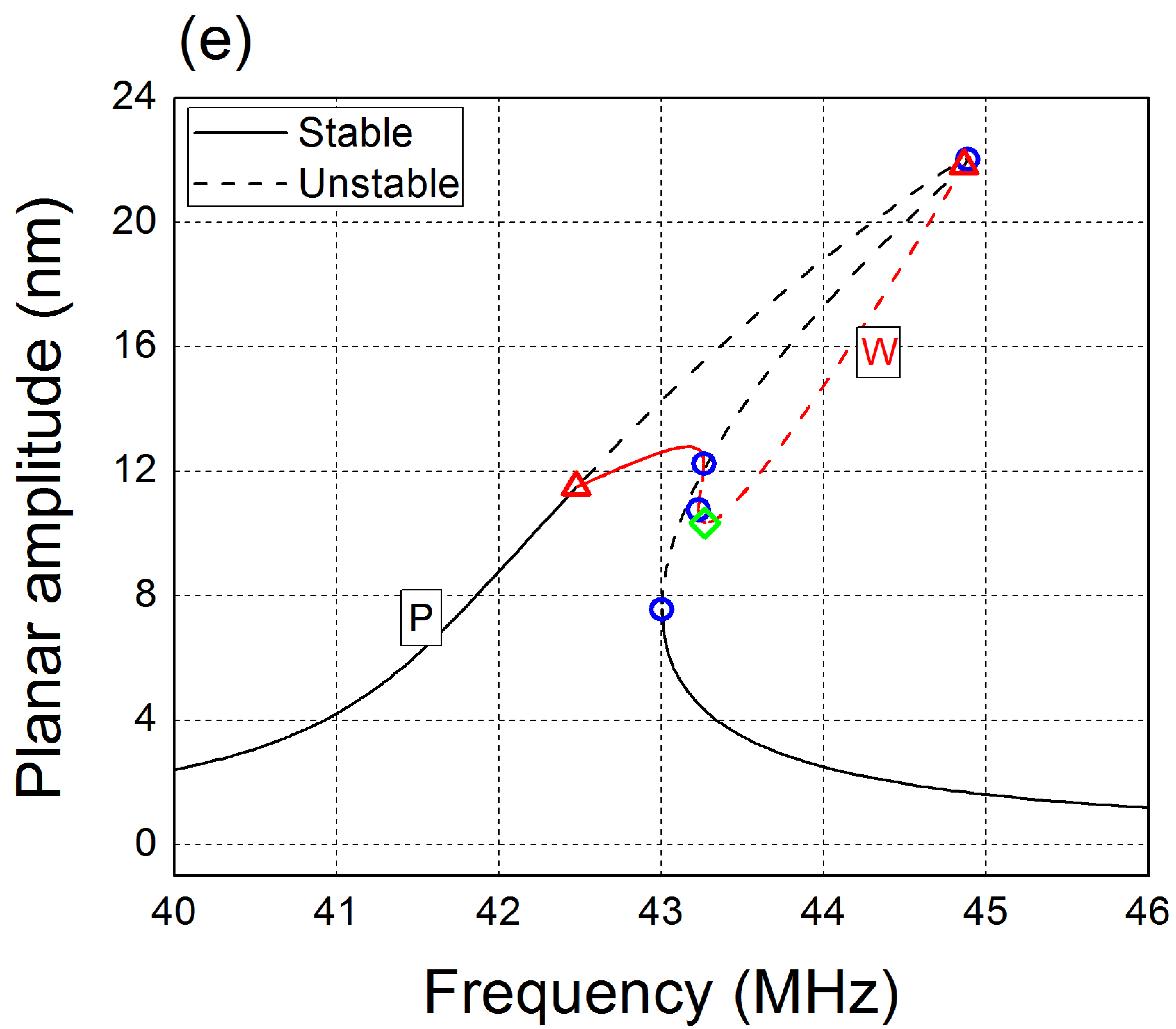}}
\hspace{0.5cm} 
\subfloat[]{\includegraphics[width=7.5cm]{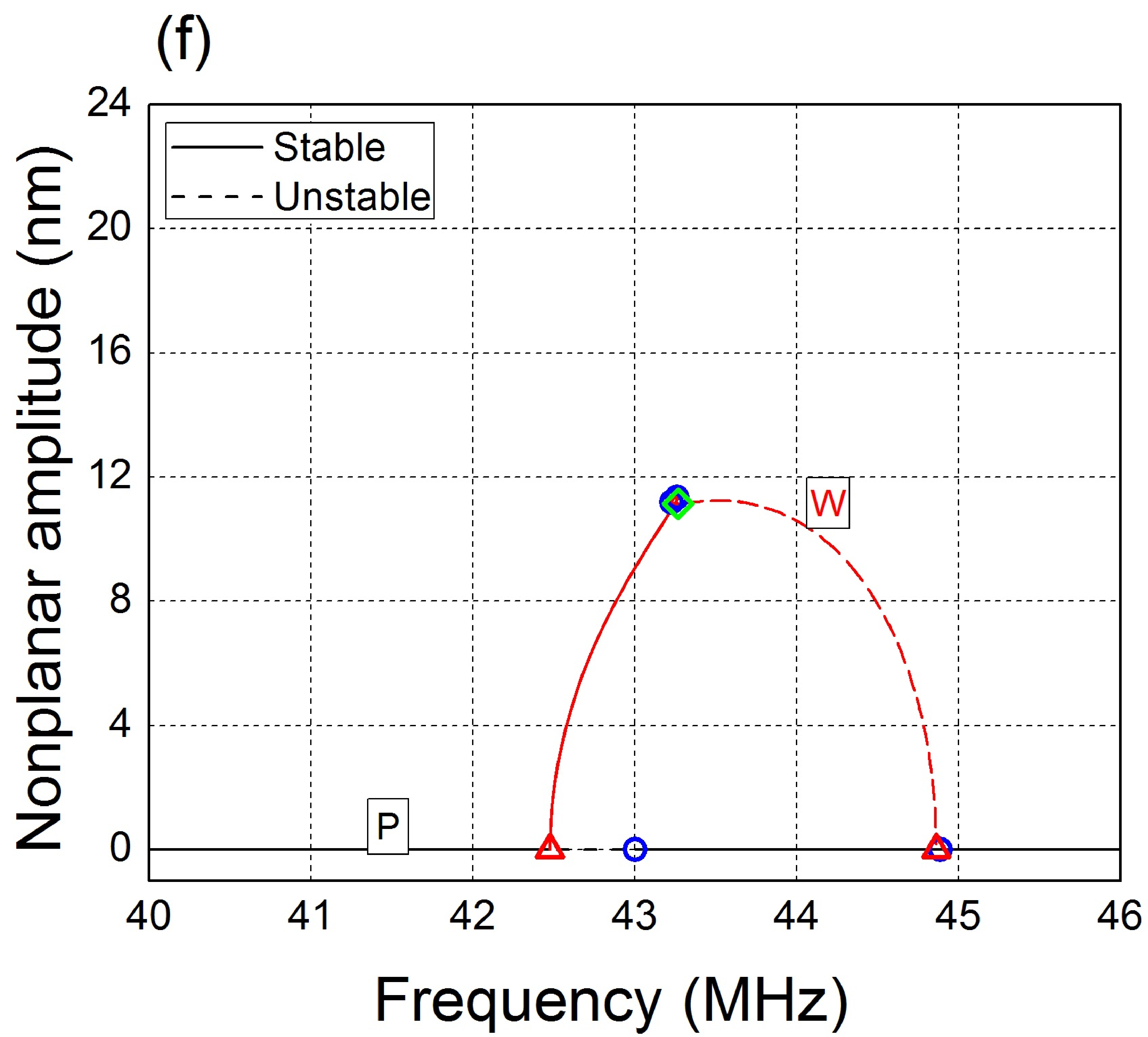}}
\caption{(a) and (b) are planar and nonplanar resonance curves, respectively, for $V_{DC}=5$ V and $V_{AC}=0.34$ V. Similarly, (c) and (d) are planar and nonplanar resonance curves for $V_{DC}=5$ V and $V_{AC}=0.38$ V; and (e) and (f) are for $V_{DC}=5$ V and $V_{AC}=0.70$ V. Unfilled circle ($\circ$), triangle ($\triangle$), and diamond ($\diamond$) denote cyclic-fold, branch, and torus bifurcation points respectively.}
\label{wfig.rc-5}
\end{figure*}
\begin{figure*}
\captionsetup[subfigure]{labelformat=empty}
\subfloat{\includegraphics[width=7.5cm]{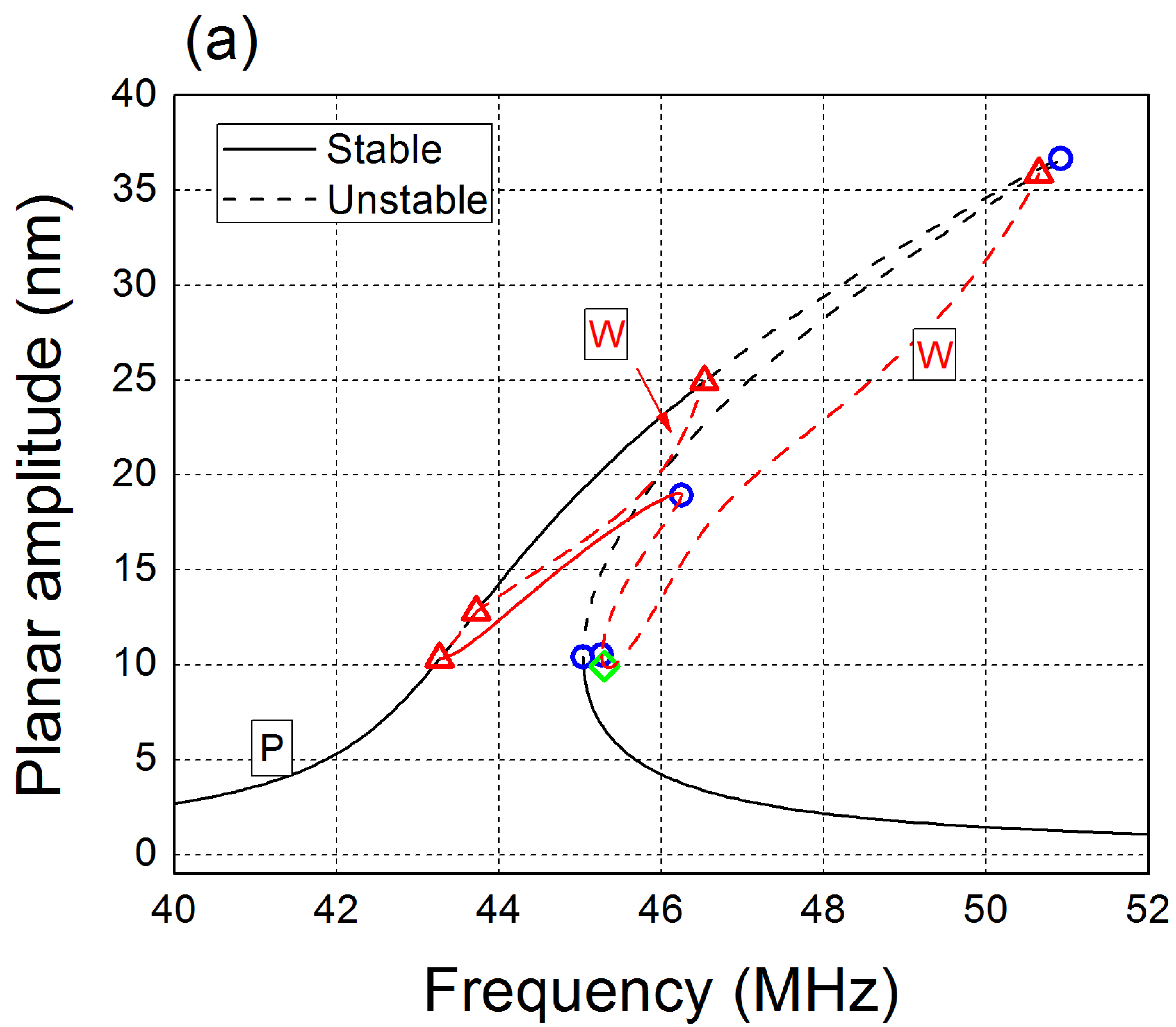}}
\hspace{0.5cm} 
\subfloat{\includegraphics[width=7.5cm]{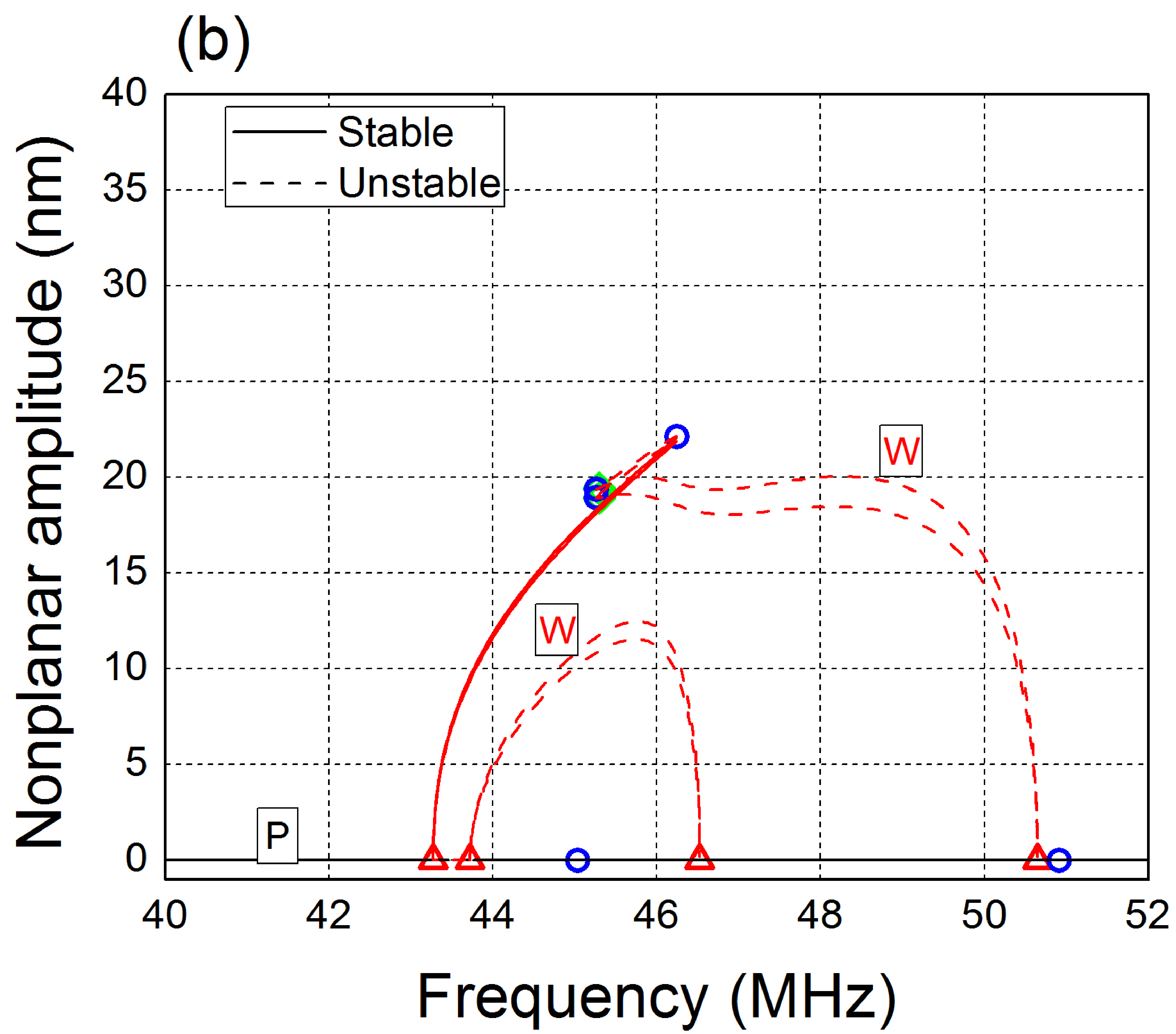}}
\caption{(a) and (b) are planar and nonplanar resonance curves, respectively,  for $V_{DC}=17.5$ V and $V_{AC}=0.40$. Unfilled circle ($\circ$), triangle ($\triangle$), and diamond ($\diamond$) denote cyclic-fold, branch, and torus bifurcation points respectively.}
\label{wfig.rc-17p5-0p4}
\end{figure*}
\begin{figure*}
\captionsetup[subfigure]{labelformat=empty}
\subfloat{\includegraphics[width=7.5cm]{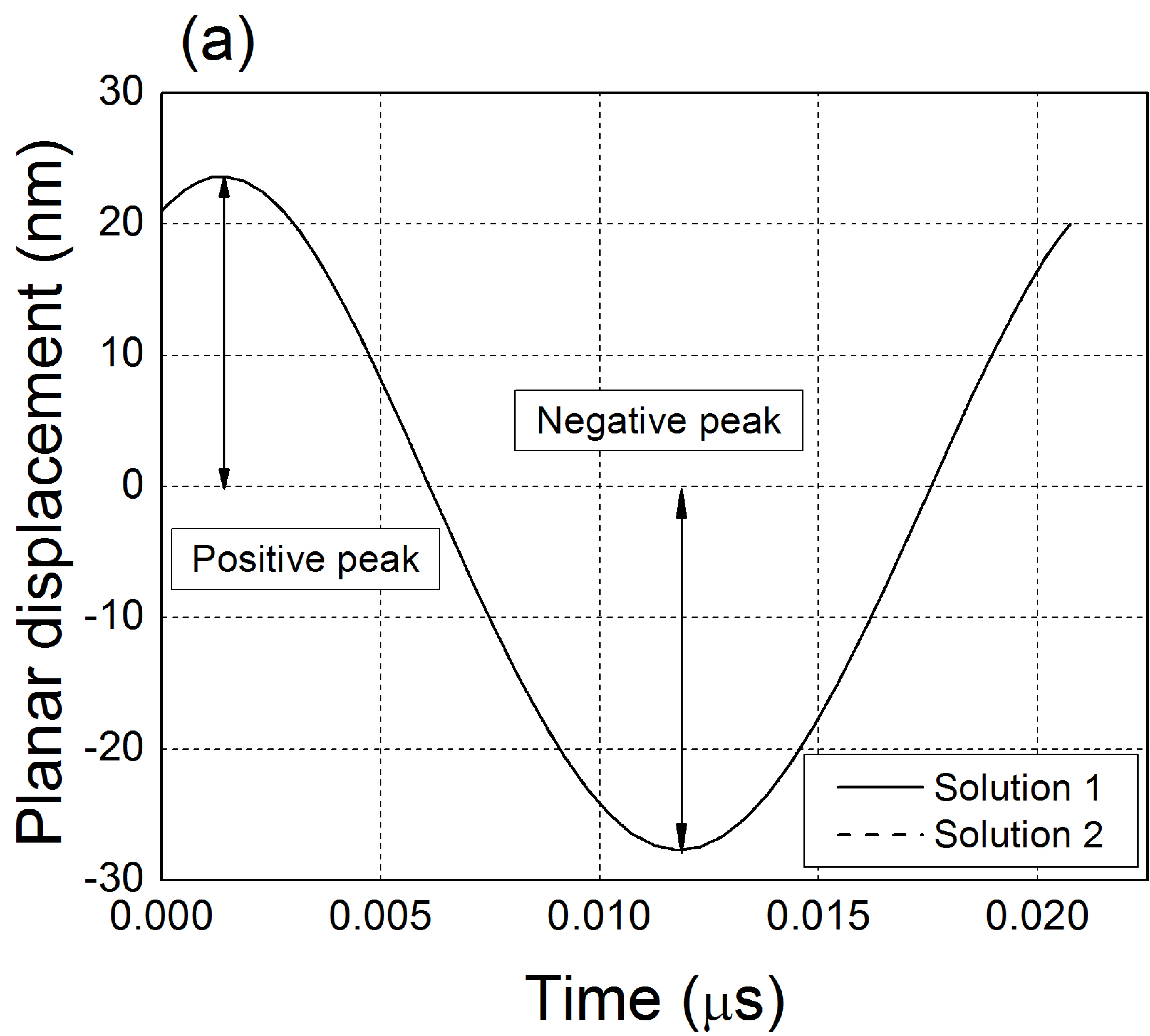}}
\hspace{0.5cm} 
\subfloat{\includegraphics[width=7.5cm]{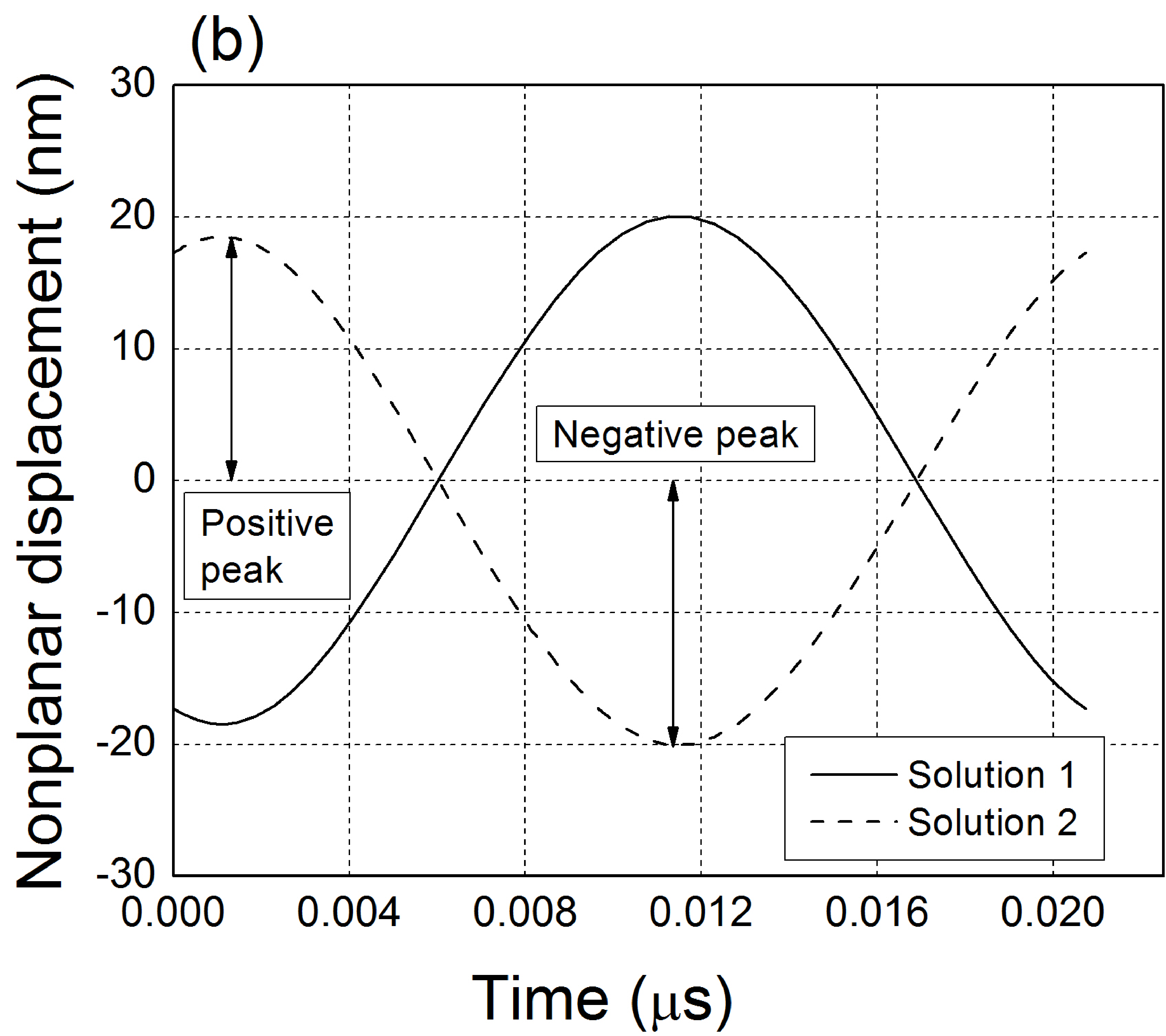}}
\caption{(a) and (b) are planar and nonplanar components, respectively, of periodic solutions for $V_{DC}=17.5$ V, $V_{AC}=0.40$ V, and $\omega_f = 48.21$ MHz.}
\label{wfig.bistability}
\end{figure*}

After setting up the governing differential equation \eqref{weq.rom2} in the previous section, now we present the numerical solutions of the equation to show the effects of DC voltage on the whirling dynamics of the nanowire oscillator. We first explain the characteristics of planar and nonplanar resonance curves by referring Figs. \ref{wfig.rc-5}(a)-\ref{wfig.rc-5}(f). Figures \ref{wfig.rc-5}(a) and \ref{wfig.rc-5}(b) show planar and nonplanar resonance curves of the nanowire oscillator when DC voltage $V_{DC}$ is equal to 5 V and amplitude of AC voltage $V_{AC}$ is equal to 0.34 V. In this case, the resonance behaviour of the nanowire oscillator is similar to the resonance curve of a Duffing oscillator; the motion of the nanowire is planar because nonplanar amplitude is zero throughout the range of forcing frequency as can be seen in Fig. \ref{wfig.rc-5}(b). Two unfilled circles are marked in the figures characterise the nonlinear nature of the resonance curves; these circles represent cyclic-fold bifurcation points \cite{nayfeh19951} in the periodic solutions where jump up and jump down behaviour can be observed in backward and forward frequency sweep respectively. However, as the magnitude of $V_{AC}$ increases to 0.38 V, W-branch emanates from P-branch at two branch bifurcation points \cite{conley2010, ermentrout2002} represented with unfilled triangles as shown in Figs. \ref{wfig.rc-5}(c) and \ref{wfig.rc-5}(d). The nanowire oscillation can be described as whirling motion for the range of forcing frequency where both planar and nonplanar amplitudes have non-zero magnitude. We have observed additional bifurcation points on resonance curves as the magnitude of $V_{AC}$ increases further to 0.70 V, and shown in Figs. \ref{wfig.rc-5}(e) and \ref{wfig.rc-5}(f). One can observe from these figures that both stable and unstable solutions exist on W-branch separated by cyclic fold bifurcation points and torus bifurcation points (represented by unfilled diamond). Similar nature of W-branch has also been observed by Abe \cite{abe2010} in the investigation of nonlinear dynamics of suspended cables.\\
\indent Upon increasing the magnitude of DC voltage to 17.5 V, we have observed qualitative distinct behaviour in resonance curves for the transition in planar to whirling motion of the nanowire oscillator. The qualitative change in behaviour can be observed by comparing planar and nonplanar resonance curves in Figs. \ref{wfig.rc-5}(e) and \ref{wfig.rc-5}(f) corresponding to $V_{DC}$ = 5 V and $V_{AC}$ = 0.70 V with Figs. \ref{wfig.rc-17p5-0p4}(a) and \ref{wfig.rc-17p5-0p4}(b) corresponding to $V_{DC}$ = 17.5 V and $V_{AC}$ = 0.40 V. There are four branch bifurcation points corresponding to two W-branches in Figs. \ref{wfig.rc-17p5-0p4}(a) and \ref{wfig.rc-17p5-0p4}(b), as compared to only two bifurcation points corresponding to one W-branch in Fig. \ref{wfig.rc-5}(e) and \ref{wfig.rc-5}(f).\\
\indent In addition to the change in qualitative nature of resonance curves, another effect of DC voltage is to bring about asymmetry in oscillation because of the presence of quadratic nonlinear terms in Eq. \eqref{weq.rom2}; this effect is demonstrated in Fig. \ref{wfig.rc-17p5-0p4}(b). As can be observed from Fig. \ref{wfig.rc-17p5-0p4}(b), there are two curves corresponding to each W-branch of the nonplanar resonance curve, and these two curves represent two periodic solutions of Eq. \eqref{weq.rom2}. This can be understood by observing that if a periodic solution $[u_{d,pr}(t)\,\,\, v_{d,pr}(t)]$ exists for Eq. \eqref{weq.rom2} then another periodic solution $[u_{d,pr}(t) \,\,\,-v_{d,pr}(t)]$ also exists because of absence of external forcing in nonplanar direction. Due to presence of quadratic nonlinearities in the system, these two periodic solutions are asymmetric with respect to positive and negative sides of oscillation. To further demonstrate this, we plot two periodic solutions (labelled as 1 and 2) in Figs. \ref{wfig.bistability}(a) and \ref{wfig.bistability}(b) for $\hat \omega_f = 48.21 $ MHz corresponding to W-branch of the resonance curve of Figs. \ref{wfig.rc-17p5-0p4}(a) and \ref{wfig.rc-17p5-0p4}(b) -- planar solution $u_{d,pr}(t)$ is the same for both figures, whereas nonplanar solutions ($v_{d,pr}(t)$ and $-v_{d,pr}(t)$) are exactly opposite to each other. Thus, difference in positive and negative peak amplitudes in the both planar and nonplanar solutions reflect the asymmetric nature of the oscillation and is due to presence of DC voltage. It may be noted that all resonance curves plotted in this paper, except in Figs. \ref{wfig.rc-17p5-0p4}(a) and \ref{wfig.rc-17p5-0p4}(b), are corresponding to average of positive and negative peak amplitudes; Figs. \ref{wfig.rc-17p5-0p4}(a) and \ref{wfig.rc-17p5-0p4}(b) are plot of positive peak amplitudes of periodic solutions with respect to frequency $\omega_f$ of AC voltage.  \\
\indent We have further investigated analytically the dynamics of the nanowire oscillator using second-order averaging method. The results are presented in following sections to quantify the effects of DC voltage on the resonance behaviour and explain the observations of this section.
\section{Averaging formulation}\label{ws.ave-form}
\begin{figure*}
\captionsetup[subfigure]{labelformat=empty}
\subfloat{\includegraphics[width=7.5cm]{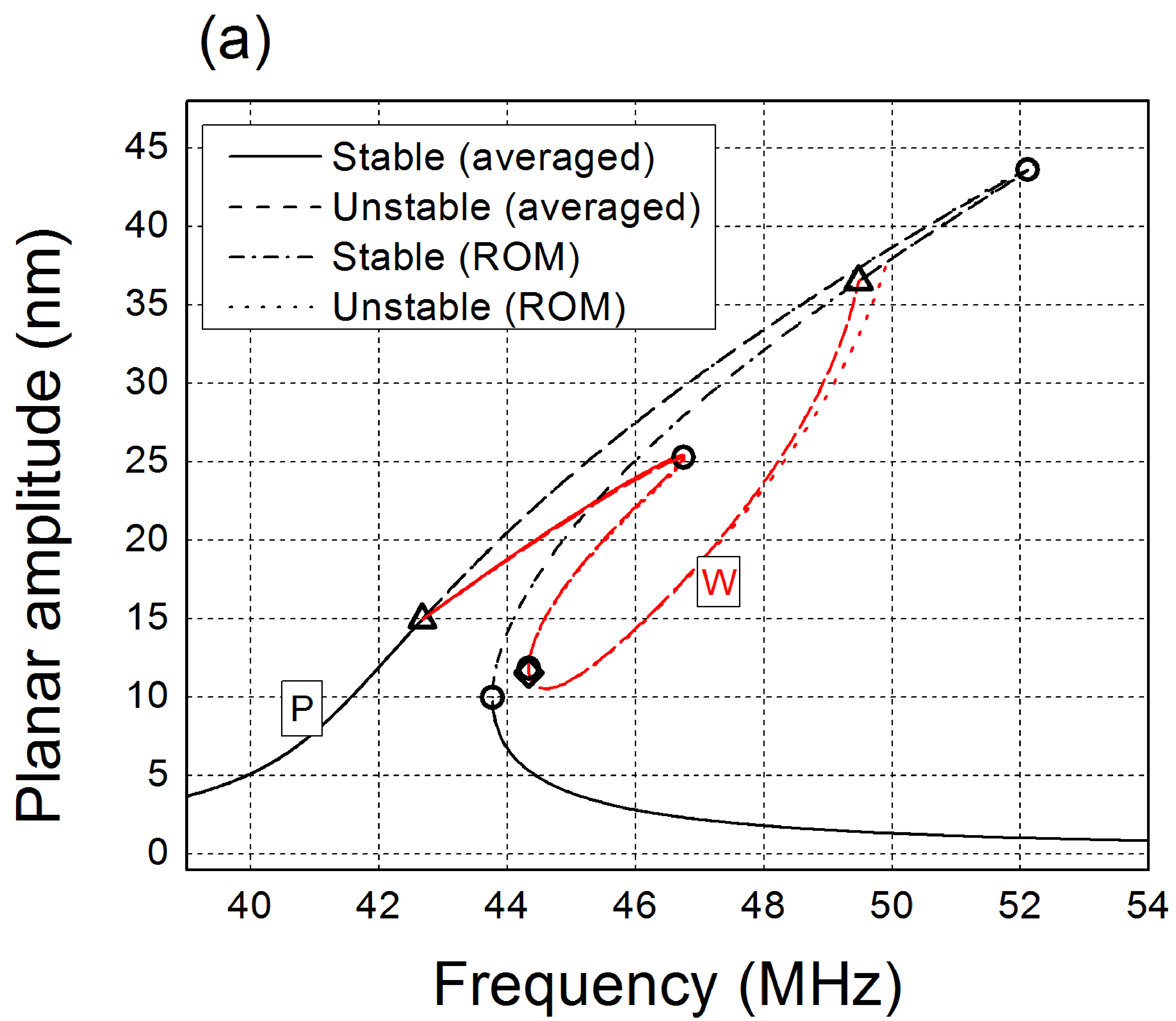}}
\hspace{0.5cm} 
\subfloat{\includegraphics[width=7.5cm]{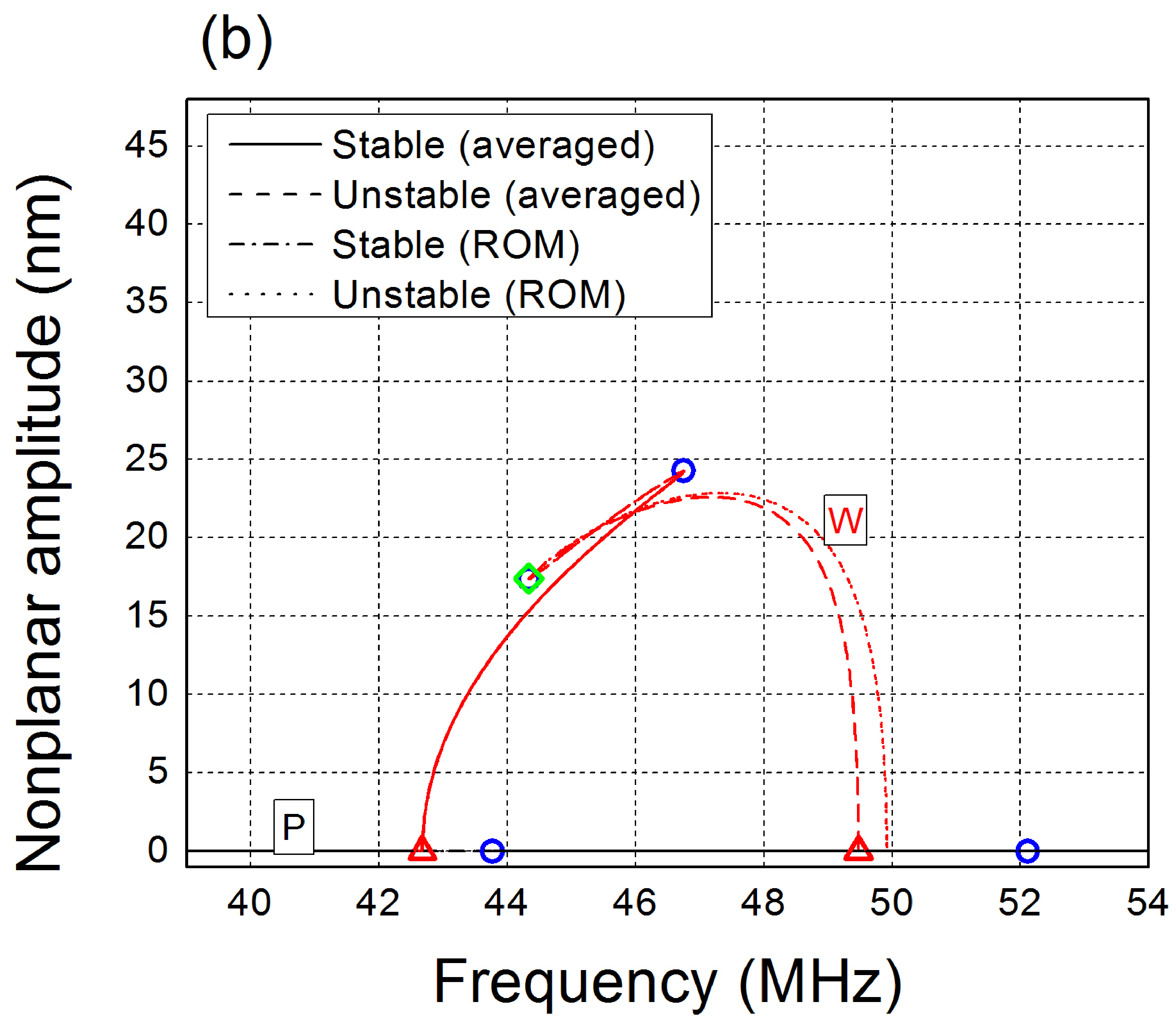}}
\caption{Comparison of resonance curves obtained by solving averaged equations \eqref{weq.averaged-eqn} and single-mode ROM \eqref{weq.rom2} for $V_{DC}=10$ V and $V_{AC}=0.80$ V: (a) planar resonance curves and (b) nonplanar resonance curves. Unfilled circle ($\circ$), triangle ($\triangle$), and diamond ($\diamond$) denote saddle-node, branch, and Hopf bifurcation points respectively.}
\label{wfig.rc-10-0p8}
\end{figure*}

Analytical investigation has been carried out by solving the whirling dynamic problem using a perturbation technique \cite{murdock1991}. Method of multiple scales and averaging method are two popular perturbation techniques frequently use for nonlinear dynamic studies \cite{cartmell, bajaj1992}. Here, we have applied averaging method for investigating the coupled oscillator problem (\ref{weq.rom2}). To solve Eq. \eqref{weq.rom2}, we assume it as a weakly nonlinear problem by rewriting coefficient of the quadratic nonlinear terms of the order $\epsilon$ and remaining terms (cubic nonlinearity, damping coefficient, and harmonic forcing) of the order $\epsilon^2$, where $\epsilon$ is a small book-keeping parameter. To analyse this problem, second-order averaging technique has been applied \cite{murdock1991}, in contrast to first-order averaging applied by Conley et al. \cite{conley2008}, for properly incorporating the effects of quadratic nonlinearity in resonance curves. It can be shown that first-order perturbation is sufficient to account for symmetric cubic nonlinearity effects on resonance curves, whereas second-order perturbation is indispensable to properly incorporate asymmetric quadratic nonlinearity \cite{nayfeh1979}. A perturbation problem has been formulated as
\begin{equation}\label{weq.pert-prob}
\begin{array}{l}
 \ddot u_d  + \omega _f^2 u_d  = \varepsilon f_{u1}^{(d)}  + \varepsilon ^2 f_{u2}^{(d)}, \\
 \ddot v_d  + \omega _f^2 v_d  = \varepsilon f_{v1}^{(d)}  + \varepsilon ^2 f_{v2}^{(d)}, \\
 f_{u1}^{(d)}  =  - k_{21} u_d^2  - k_{22} v_d^2, \\ f_{u2}^{(d)}  =  - \mu \dot u_d  - k_{31} u_d^3  - k_{312} u_d v_d^2  + p\cos (\omega _f t) - \Delta _1 u_d,  \\
 f_{v1}^{(d)}  =  - k_{212} u_d v_d, \\ f_{v2}^{(d)}  = - \mu \dot v_d  - k_{32} v_d^3  - k_{312} u_d^2 v_d  - \Delta _2 v_d \cdot  \\
\end{array}
\end{equation}
The set of coefficients of Eq. \eqref{weq.pert-prob} are related with coefficients of Eq. \eqref{weq.rom2} through a small parameter $\varepsilon_0$ as
\begin{equation*}
\begin{array}{l}
 c = \varepsilon _0^2 \mu ,\,\,k_{1u}  = \omega _u^2 ,\,\,k_{2u}  = \varepsilon _0 k_{21} ,\,\,k_{2v}  = \varepsilon _0 k_{22} ,\,\,k_{3u}  = \varepsilon _0^2 k_{31} , \\
 k_{1v}  = \omega _v^2 ,\,\,k_{2uv}  = \varepsilon _0 k_{212} ,\,\,k_{3v}  = \varepsilon _0^2 k_{32} ,\,\,k_{3uv}  = \varepsilon _0^2 \,k_{312} , \\
 2V_{DC} V_{AC} C_0  = \varepsilon _0^2 p,\,\,\omega _u^2  = \omega _f^2  + \varepsilon _0^2 \Delta _1 ,\,\,\omega _v^2  = \omega _f^2  + \varepsilon _0^2 \Delta _2 \cdot \\
 \end{array}
\end{equation*}
The solution of Eq. \eqref{weq.pert-prob} becomes solution of Eq. \eqref{weq.rom2} when $\varepsilon_0$ is equal to $\varepsilon$. In Eq. \eqref{weq.pert-prob}, two detuning parameters $\Delta_1$ and $\Delta_2$ are introduced which measure the difference of forcing frequency $\omega_f$ from first planar natural frequency $\omega_{u} = \sqrt{k_{1u}}$ and nonplanar natural frequency $\omega_{v} = \sqrt{k_{1v}}$, respectively.\\
\indent To make the perturbation problem suitable for averaging, we first transform Eq. \eqref{weq.pert-prob} in periodic standard form by expressing the displacement and velocity components as $[u_d\,\, \dot u_d]^T = \textbf{A} [x_1\,\,x_2]^T$ and $[v_d\,\, \dot v_d]^T = \textbf{A} [x_3\,\,x_4]^T$, where $\textbf{A}$ is a matrix defined as
\begin{equation*}
\textbf{A} = \left[ {\begin{array}{*{20}c}
   {\cos (\omega _f t)} & {\sin (\omega _f t)}  \\
   { - \omega _f \sin (\omega _f t)} & {\omega _f \cos (\omega _f t)}\\
\end{array}} \right]\cdot
\end{equation*}
The periodic standard form of the coupled oscillator is
\begin{equation}\label{weq.periodic-standard}
\begin{array}{l}
\displaystyle
 \frac{{d{\bf{x}}}}{{d\tau }} = \varepsilon \left( {{\bf{f}}_1  + \varepsilon {\bf{f}}_2 } \right),\,\,\,{\bf{x}} = \left[ {x_1 \,\,x_2 \,\,x_3 \,\,x_4 } \right]^T,\,\,\tau  = \omega _f t, \\
 {\bf{f}}_1  = \frac{1}{{\omega _f^2 }}\left[ \begin{array}{l}
  - f_{u1}^{(d)} \sin (\tau ) \\
 f_{u1}^{(d)} \cos (\tau ) \\
  - f_{v1}^{(d)} \sin (\tau ) \\
 f_{v1}^{(d)} \cos (\tau ) \\
 \end{array} \right],\,{\bf{f}}_2  = \frac{1}{{\omega _f^2 }}\left[ \begin{array}{l}
  - f_{u2}^{(d)} \sin (\tau ) \\
 f_{u2}^{(d)} \cos (\tau ) \\
  - f_{v2}^{(d)} \sin (\tau ) \\
 f_{v2}^{(d)} \cos (\tau ) \\
 \end{array} \right]\, \cdot  \\
 \end{array}
\end{equation}
The central idea of second-order averaging is to obtain a near identity transformation ${\bf{x}}(\tau) = {\bf{y}}(\tau) + \varepsilon {\bf{b}_1}  ({\bf{y}},\tau) + ... $, such that the averaged equation ${d\bf{y}}/d \tau = \varepsilon {\bf{g}}_1 ({\bf{y}}) + \varepsilon ^2 {\bf{g}}_2 ({\bf{y}})+ ...$ is an autonomous system \cite{murdock1991}. We can calculate $\bf{g}_1 ({\bf{y}})$ and ${\bf{b}_1}  ({\bf{y}},\tau)$ from first averaging using following expressions
\begin{equation} \label{weq.first-avg}
\begin{array}{l}
{\bf{g}}_1 ({\bf{y}}) = \frac{1}{{2\pi }}\int\limits_0^{2\pi } {{\bf{f}}_1 ({\bf{y}},\lambda )} d\lambda \,\,\,{\rm{and }} \\ {\bf{b}_1}({\bf{y}},\tau ) = \int\limits_0^\tau  {\left( {{\bf{f}}_1 ({\bf{y}},\lambda ) - {\bf{g}}_1 ({\bf{y}})} \right)} d\lambda \cdot
\end{array}
\end{equation}
Further, we have computed $\bf{g}_2 ({\bf{y}})$ using second averaging
\begin{equation} \label{weq.second-avg}
\begin{array}{l}
\displaystyle {\bf{g}}_2 ({\bf{y}}) = \frac{1}{{2\pi }}\cdot \\ \displaystyle \int\limits_0^{2\pi } {\left( {{\bf{f}}_2 ({\bf{y}},\lambda ) + \left. {\frac{{\partial {\bf{f}}_1 ({\bf{x}},\lambda )}}{{\partial {\bf{x}}}}} \right|_{{\bf{x}} = {\bf{y}}} {\bf{b}}_1 ({\bf{y}},\lambda ) - \frac{{\partial {\bf{b}}_1 ({\bf{y}},\lambda )}}{{\partial {\bf{y}}}}{\bf{g}}_1 ({\bf{y}})} \right)} d\lambda \cdot
\end{array}
\end{equation}
After computing Eqs. \eqref{weq.first-avg} and \eqref{weq.second-avg}, we have obtained the averaged equations which determine amplitude and phase evolution of the nanowire oscillator in time coordinate $t$ as
\begin{equation} \label{weq.averaged-eqn}
\begin{array}{l}
 \dot y_1  =  - \mu _0 y_1  + \left( {\Omega _1  + \gamma _1 ( y_1^2  + y_2^2)  + \gamma _3 ( y_3^2  + y_4^2)  } \right)y_2  + \\ \gamma _4 y_3 (y_1 y_4  - y_2 y_3),  \\
 \dot y_2  =  - \mu _0 y_2  - \left( {\Omega _1  + \gamma _1 ( y_1^2  + y_2^2)  + \gamma _3 ( y_3^2  + y_4^2) } \right)y_1  +  \\ \gamma _4 y_4 (y_1 y_4  - y_2 y_3)  + p_0, \\
 \dot y_3  =  - \mu _0 y_3  + \left( {\Omega _2  + \gamma _2 ( y_3^2  + y_4^2)  + \gamma _3 ( y_1^2  + y_2^2)  } \right)y_4  - \\ \gamma _4 y_1 (y_1 y_4  - y_2 y_3) , \\
 \dot y_4  =  - \mu _0 y_4  - \left( {\Omega _2  + \gamma _2 ( y_3^2  + y_4^2)  + \gamma _3 ( y_1^2  + y_2^2)  } \right)y_3  - \\ \gamma _4 y_2 (y_1 y_4  - y_2 y_3) , \\
 \end{array}
\end{equation}
where coefficients are
\begin{equation}\label{weq.nonlinear-parameter2}
\displaystyle
\begin{array}{l}
\displaystyle
 \mu _0  = \frac{c}{2},\,\,p_0  = \frac{{2 V_{DC} V_{AC}C_0}}{{2\omega _f }},\,\,\Omega _1  = \frac{{\omega _u^2  - \omega _f^2 }}{{2\omega _f }},\,\,\Omega _2  = \frac{{\omega _v^2  - \omega _f^2 }}{{2\omega _f }}, \\
 \displaystyle
 \gamma_1  = \frac{{3k_{3u} }}{{8\omega _f }} - \frac{{5k_{2u}^2 }}{{12\omega _f^3 }},\,\,\gamma _3  = \frac{{3k_{3uv} }}{{8\omega _f }} - \frac{{5k_{2uv} k_{2u} }}{{24\omega _f^3 }} - \frac{{5k_{2uv} k_{2v} }}{{12\omega _f^3 }}, \\
 \displaystyle
 \gamma_2  = \frac{{3k_{3v} }}{{8\omega _f }} - \frac{{5k_{2v}^2 }}{{12\omega _f^3 }},\,\,\gamma _4  = \frac{{k_{3uv} }}{{4\omega _f }} + \frac{{k_{2uv} k_{2u} }}{{12\omega _f^3 }} - \frac{{k_{2uv} k_{2v} }}{{2\omega _f^3 }} \cdot\\
 \end{array}
\end{equation}
In Eq. \eqref{weq.nonlinear-parameter2}, the detuning parameters $\Omega_1$ and $\Omega_2$ are modified form of $\Delta_1$ and $\Delta_2$ of Eq. \eqref{weq.pert-prob} respectively. Steady state solutions or equilibrium points of Eq. \eqref{weq.averaged-eqn} provide periodic solution of the coupled oscillator (\ref{weq.rom2}). As can be verified by comparing Eq. \eqref{weq.averaged-eqn} with the formulation of Bajaj and Johnson \cite{bajaj1992},  Equation \eqref{weq.averaged-eqn} can be reduced to averaged equation of symmetric coupled oscillator problem of sinusoidally forced string by setting  $\omega_{u}$ is equal to $\omega_{v}$ and neglecting the effect of quadratic nonlinearities in Eq. \eqref{weq.nonlinear-parameter2}. We have also derived expressions for planar $u_d (t)$ and nonplanar $v_d (t)$ amplitude after computation of second integral of Eq. \eqref{weq.first-avg} as
\begin{equation}\label{weq.per-soln}
\begin{array}{l}
\displaystyle  u_d (t) = r_u \cos (\omega _f t + \theta _u ) - \left( {\frac{{k_{2u} }}{{2\omega _f^2 }}r_u^2  + \frac{{k_{2v} }}{{2\omega _f^2 }}r_v^2 } \right) + \\ \displaystyle \frac{{k_{2u} }}{{6\omega _f^2 }}r_u^2 \cos (2\omega _f t + 2\theta _u ) + \frac{{k_{2v} }}{{6\omega _f^2 }}r_v^2 \cos (2\omega _f t + 2\theta _v ), \\
\displaystyle v_d (t) = r_v \cos (\omega _f t + \theta _v ) - \frac{{k_{2uv} }}{{2\omega _f^2 }}\cos (\theta _u  - \theta _v )r_u r_v  + \\ \displaystyle \frac{{k_{2uv} }}{{6\omega _f^2 }}r_u r_v \cos (2\omega _f t + \theta _u  + \theta _v ) \cdot
 \end{array}
\end{equation}
Here $r_u=\sqrt{y_1^2  + y_2^2}$ and $r_v=\sqrt{y_3^2  + y_4^2}$ are planar and nonplanar amplitudes respectively, and can be obtained by substituting $y_1  = r_u \cos (\theta _u ),\,\, y_2  =  - r_u \sin (\theta _u ),\,\, y_3  = r_v \cos (\theta _v )$, and $y_4  =  - r_v \sin (\theta _v )$.\\
\indent We can easily infer the quantitative effects of DC voltage on whirling dynamics of the nanowire oscillator by observing expressions for parameters $\gamma_1,\,\gamma_2,\,\gamma_3,\,\gamma_4,\,\Omega_1,\,\text{and}\,\Omega_2$ in Eq. \eqref{weq.nonlinear-parameter2} and the coefficients of nonlinear terms in Eq. \eqref{weq.nonlinear-parameter1}. The magnitude of DC voltage $V_{DC}$ modifies these parameters, ultimately affecting the dynamics of the nanowire oscillator; Table \ref{wtab.nonlinear} shows the variation of $\gamma_1$, $\gamma_2$, $\gamma_3$, and $\gamma_4$ with respect to the magnitude of $V_{DC}$. In Eq. \eqref{weq.averaged-eqn}, $\gamma_1$ is an important parameter which decides the hardening/softening nature of the planar resonance curves \cite{bhushan2013}; the planar resonance curves display hardening behaviour for positive value of $\gamma_1$ and softening for negative value. As can be seen from the table, the value of $\gamma_1$ decreases with increment of $V_{DC}$, reduction of about 49 \% with rise of $V_{DC}$ from 5 V to 17.5 V. Moreover, other parameters $\gamma_2$, $\gamma_3$ and $\gamma_4$, which also govern whirling dynamics of the nanowire oscillator, are also significantly affected by the change in magnitude of DC voltage (refer Table \ref{wtab.nonlinear}).\\
\indent We have discussed the asymmetric nature of dynamic responses of the nanowire oscillator in Section \ref{ws.dc-voltage}. This asymmetric nature can be easily inferred from expressions for $u_d(t)$ and $v_d(t)$ in (\ref{weq.per-soln}) where additional constant components are present apart from harmonic terms. These constant terms ${ - \left( {k_{2u} r_u^2  + k_{2v} r_v^2 } \right)/2\omega _f^2 }$ and ${-\left( {k_{2uv} \cos (\theta _u  - \theta _v )r_u r_v } \right)/2\omega _f^2 }$ in planar and nonplanar solutions, respectively, arise due to presence of quadratic nonlinearities, and make overall responses of  $u_d(t)$ and $v_d(t)$ asymmetric.\\
\begin{table}
\caption{Values of nonlinear parameters for different magnitudes of $V_{DC}$ at $\omega_f = 20$.}
\label{wtab.nonlinear}
\begin{center}
\begin{tabular}{|c|c|c|c|}
  \hline
  $V_{DC}$&             5.0 V                &               10 V                &     17.5 V   \\
  \hline
  $\gamma_1$  &             814.556               &              773.722                 &     418.096   \\
  \hline
  $\gamma_2$  &           817.009               &                 812.391               &   771.916\\
  \hline
  $\gamma_3$  &               815.803              &               792.880                &   592.093      \\
  \hline
  $\gamma_4$  &            544.506              &                538.899             &  489.693  \\
  \hline
\end{tabular}
\end{center}
\end{table}
\indent We have solved averaged equations (\ref{weq.averaged-eqn}) using XPPAUT to obtain planar and nonplanar resonance curves for $V_{DC} = 10$ V and $V_{AC} = 0.80$ V. As can be seen Figs. \ref{wfig.rc-10-0p8}(a) and \ref{wfig.rc-10-0p8}(b), the averaging solutions are in good agreement with the solutions of single-mode ROM (\ref{weq.rom2}); however, there is small error in the perturbation solution for larger values of detuning parameters $\Omega_1$ and $\Omega_2$. Various bifurcation points in the averaging solutions are depicted as unfilled circles (saddle-node bifurcation), triangles (pitchfork or branch bifurcation), and diamonds (Hopf bifurcation) \cite{ermentrout2002, nayfeh19951, bajaj1992}. So, we can conclude that the averaged equations are fairly capable to capture the whirling dynamics of the nanowire oscillator. We further discuss the qualitative change in the whirling dynamics of the nanowire oscillator with variation of the magnitude of DC voltage using the averaged equations (\ref{weq.averaged-eqn}) in next section.
\section{Understanding the planar to whirling motion transition} \label{ws.ptow}
\begin{figure*}
\captionsetup[subfigure]{labelformat=empty}
\subfloat{\includegraphics[width=7.5cm]{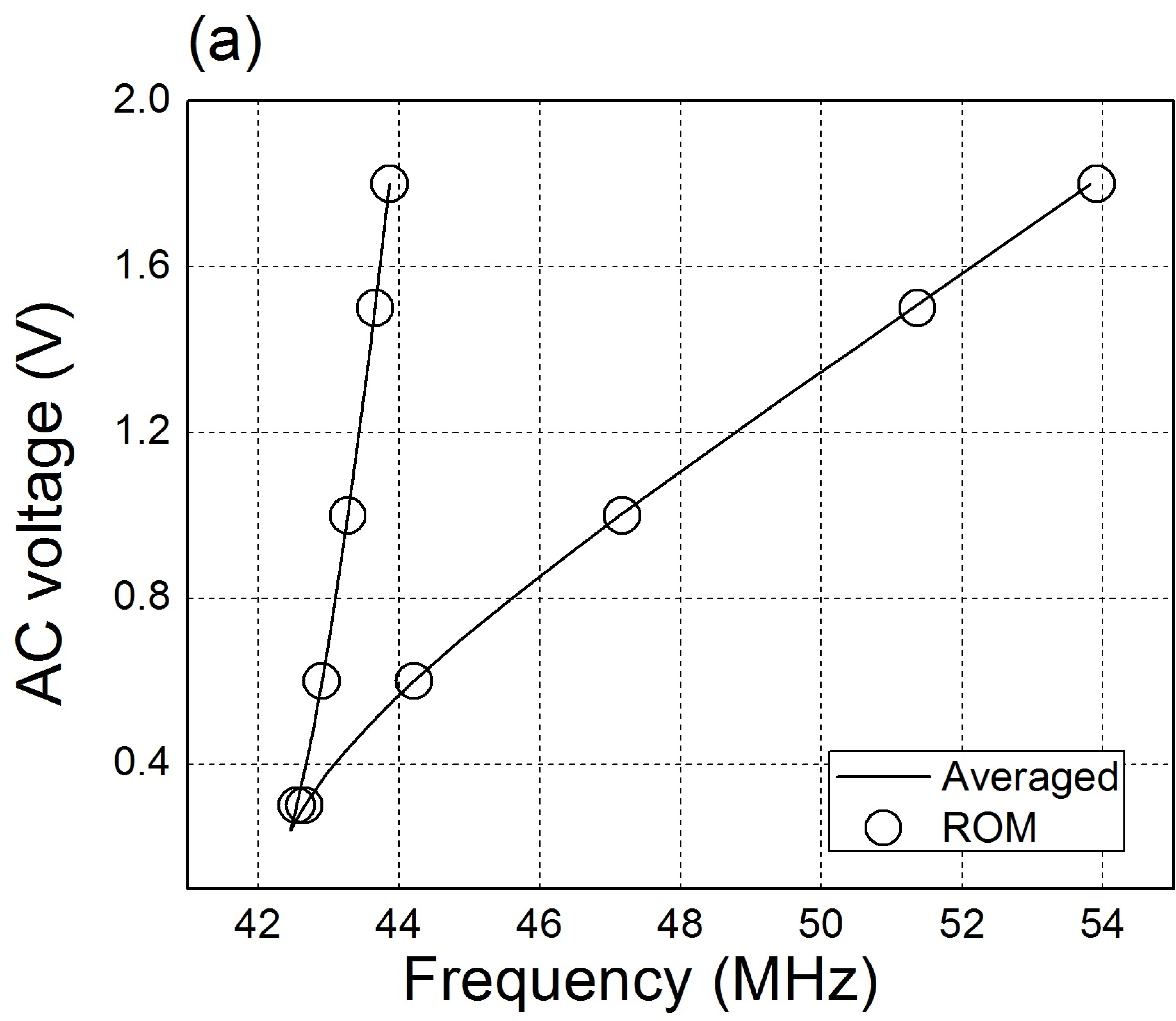}}
\hspace{0.5cm} 
\subfloat{\includegraphics[width=7.5cm]{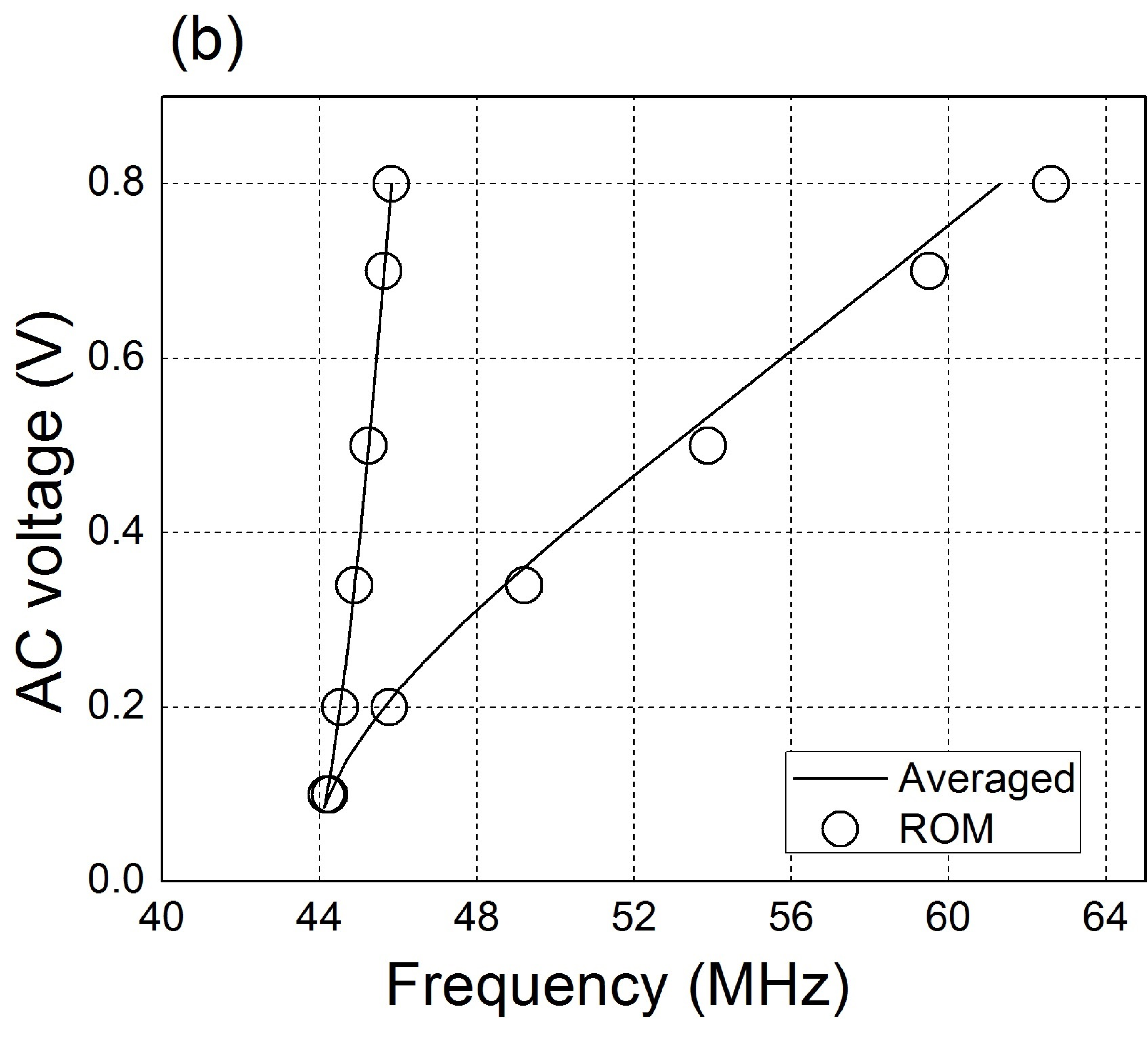}}
\caption{Saddle-node bifurcation in AC voltage - forcing frequency ($V_{AC}- \omega_f$) plane for (a) $V_{DC}=5$ V and (b) $V_{DC} = 17.5$ V.}
\label{wfig.bifs}
\end{figure*}

\begin{figure*}
\captionsetup[subfigure]{labelformat=empty}
\subfloat{\includegraphics[width=7.5cm]{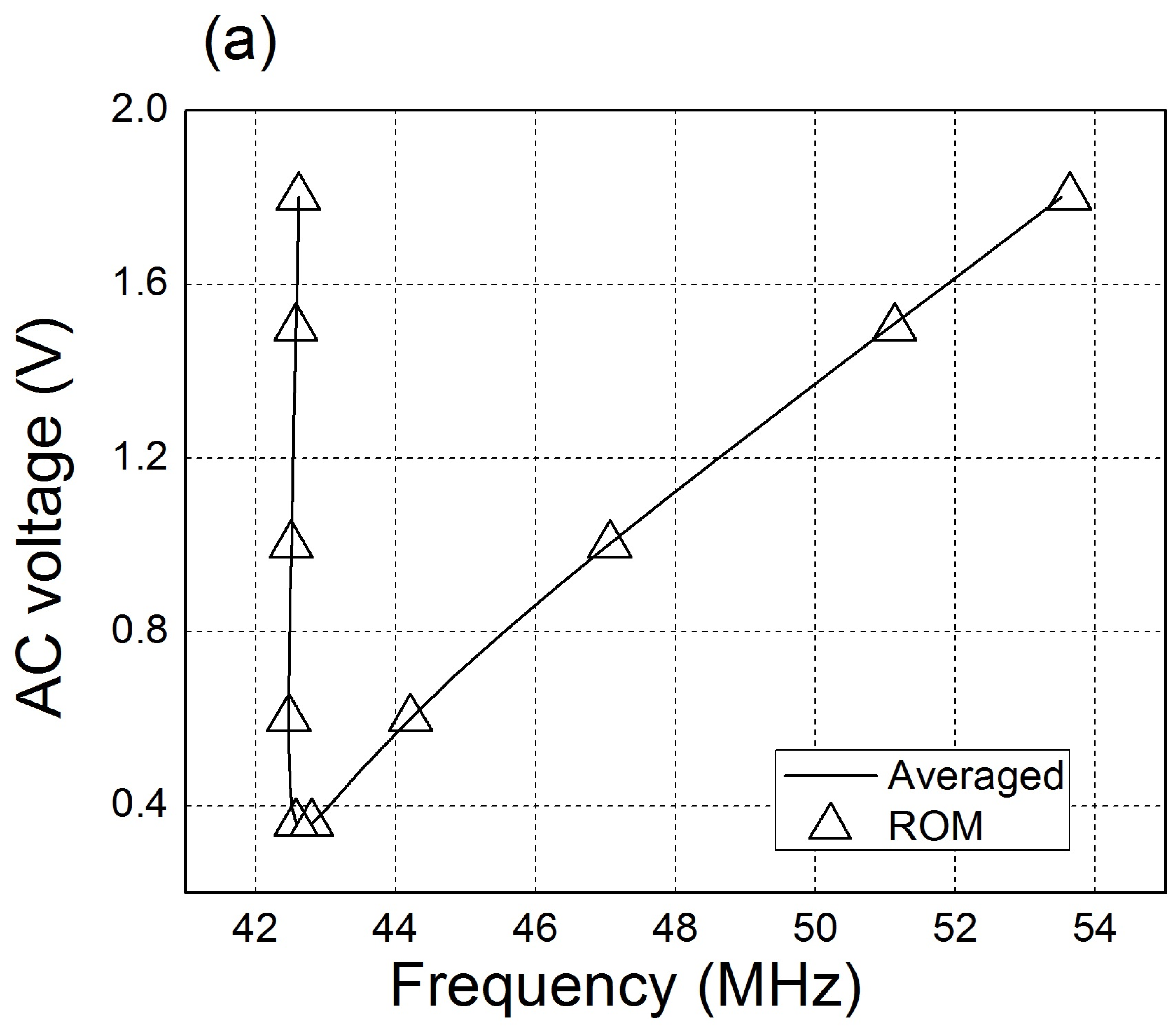}}
\hspace{0.5cm} 
\subfloat{\includegraphics[width=7.5cm]{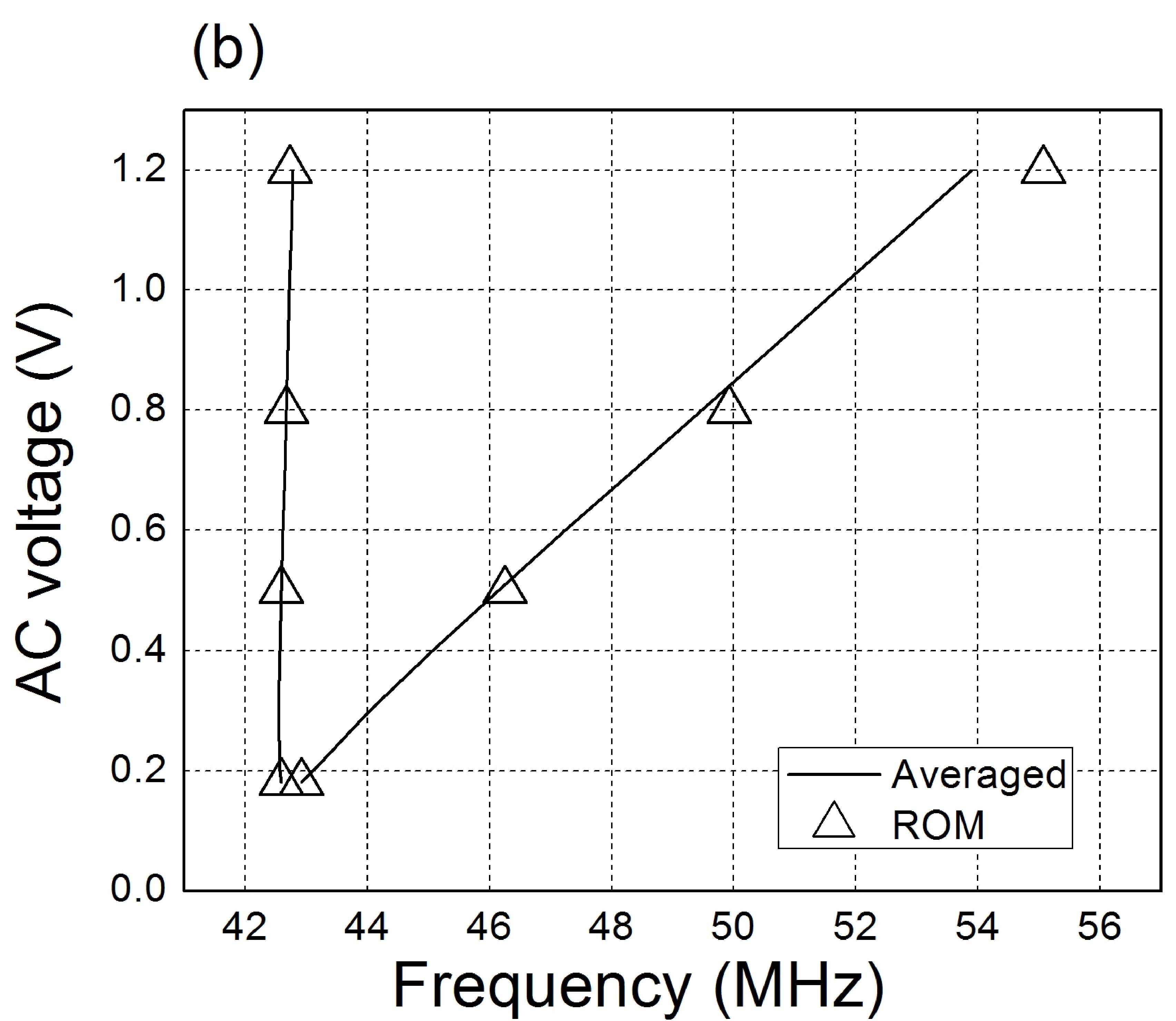}}\\
\subfloat{\includegraphics[width=7.5cm]{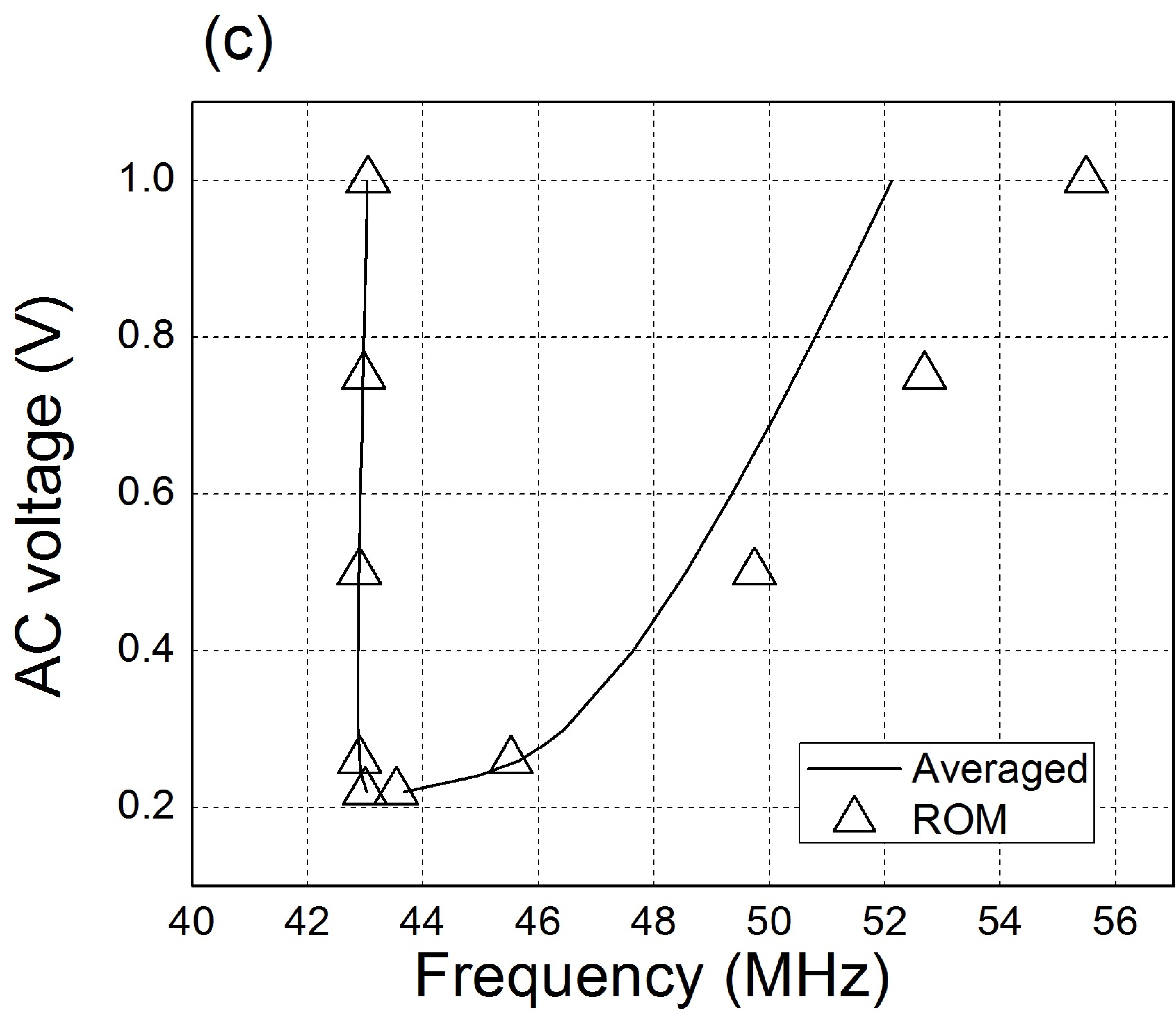}}
\hspace{0.5cm} 
\subfloat{\includegraphics[width=7.5cm]{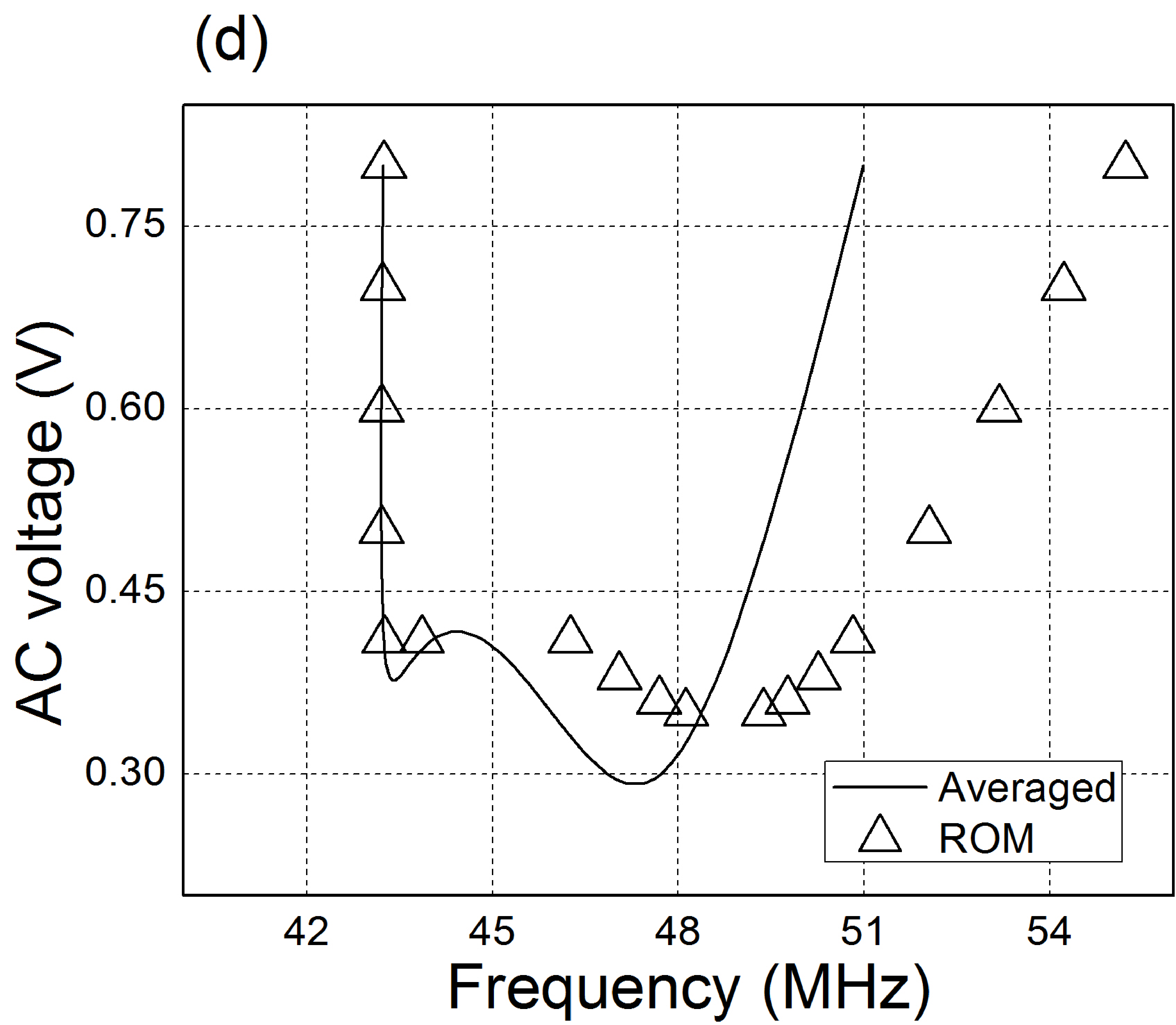}}
\caption{Branch bifurcation in AC voltage - forcing frequency ($V_{AC}- \omega_f$) plane for (a) $V_{DC}=5$ V, (b) $V_{DC}=10$ V, (c) $V_{DC}=15$ V, and (d) $V_{DC} = 17.5$ V.}
\label{wfig.bifw}
\end{figure*}

\begin{figure*}
\captionsetup[subfigure]{labelformat=empty}
\subfloat{\includegraphics[width=7.5cm]{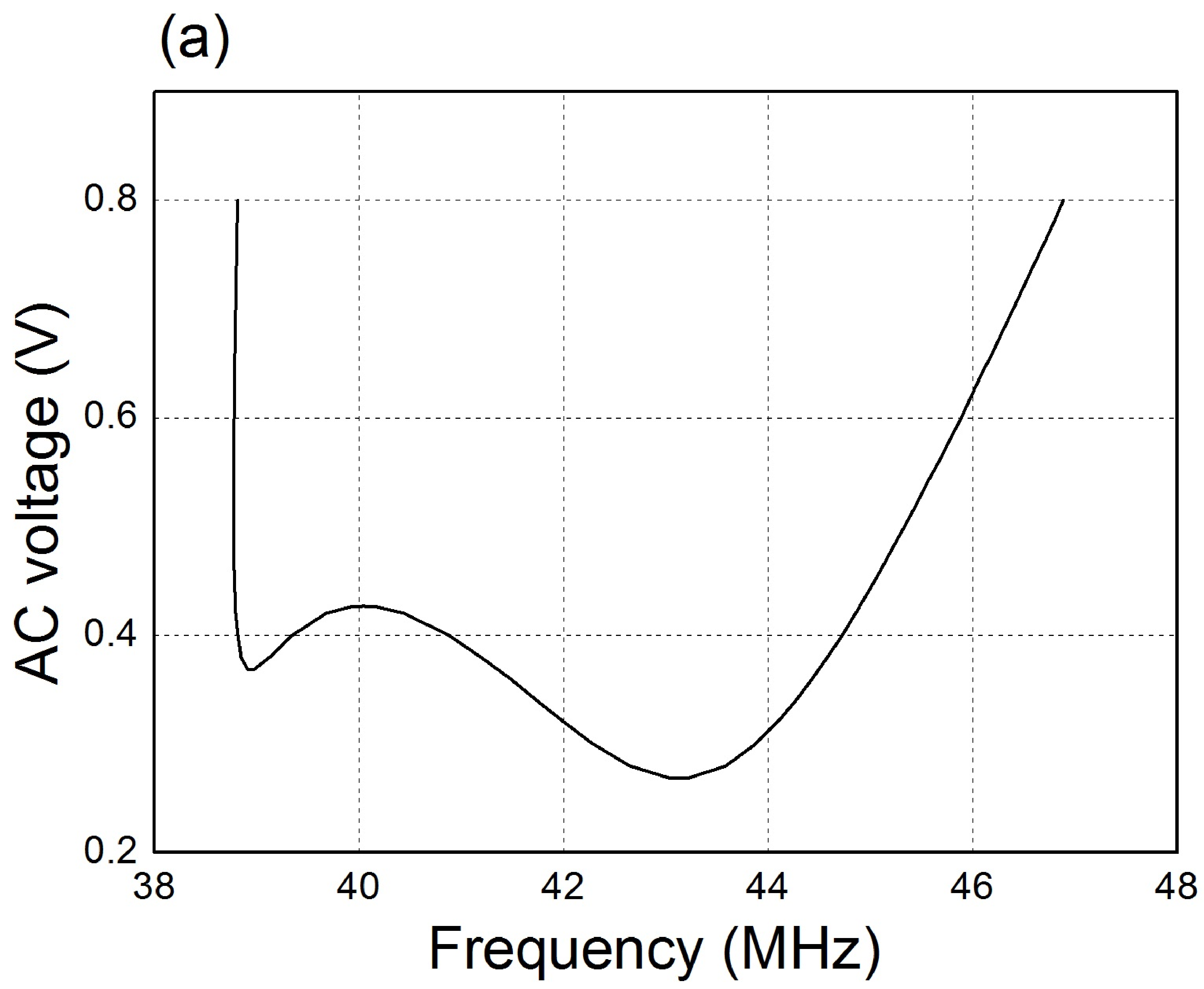}}
\hspace{0.5cm} 
\subfloat{\includegraphics[width=7.5cm]{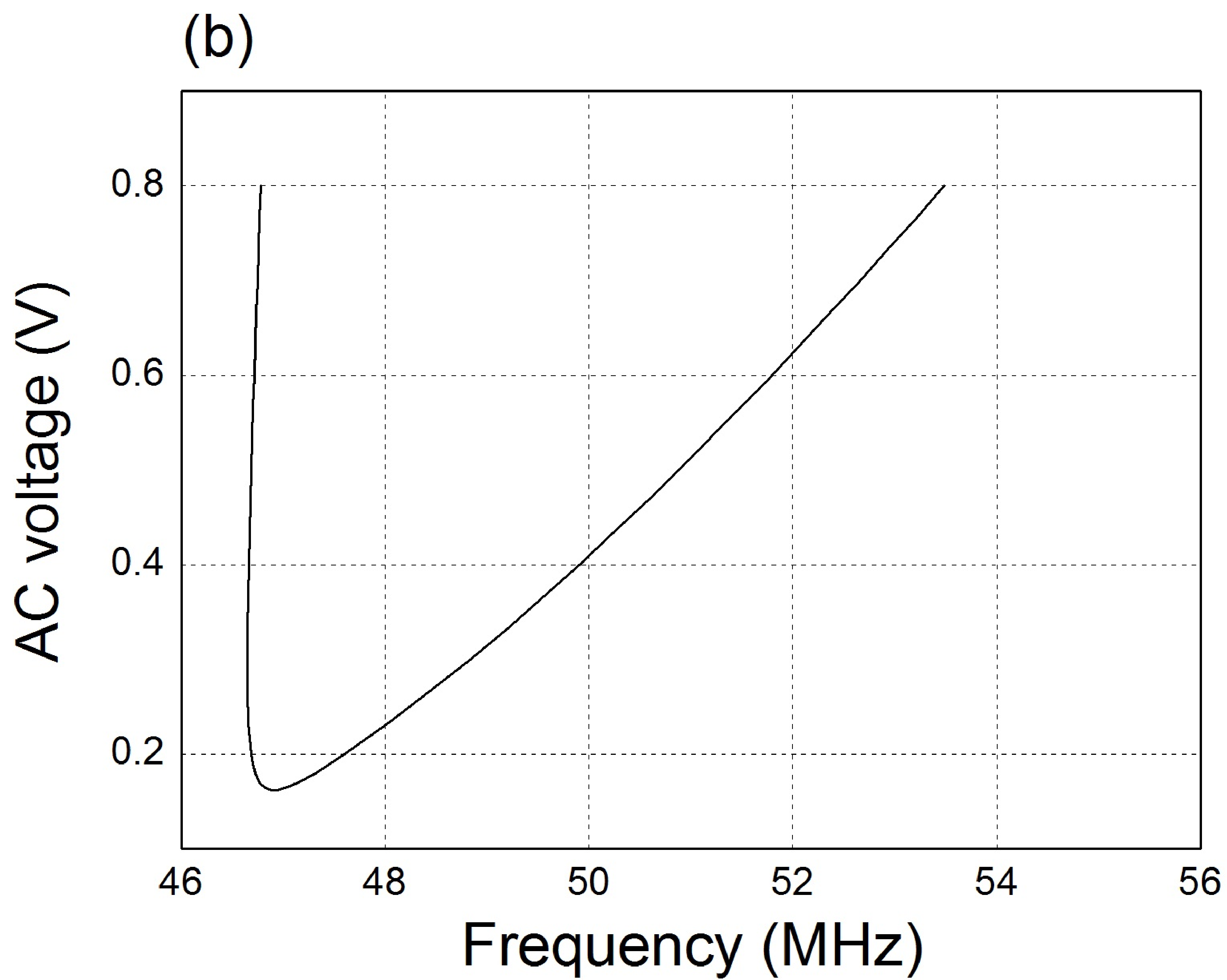}}\\
\subfloat{\includegraphics[width=7.5cm]{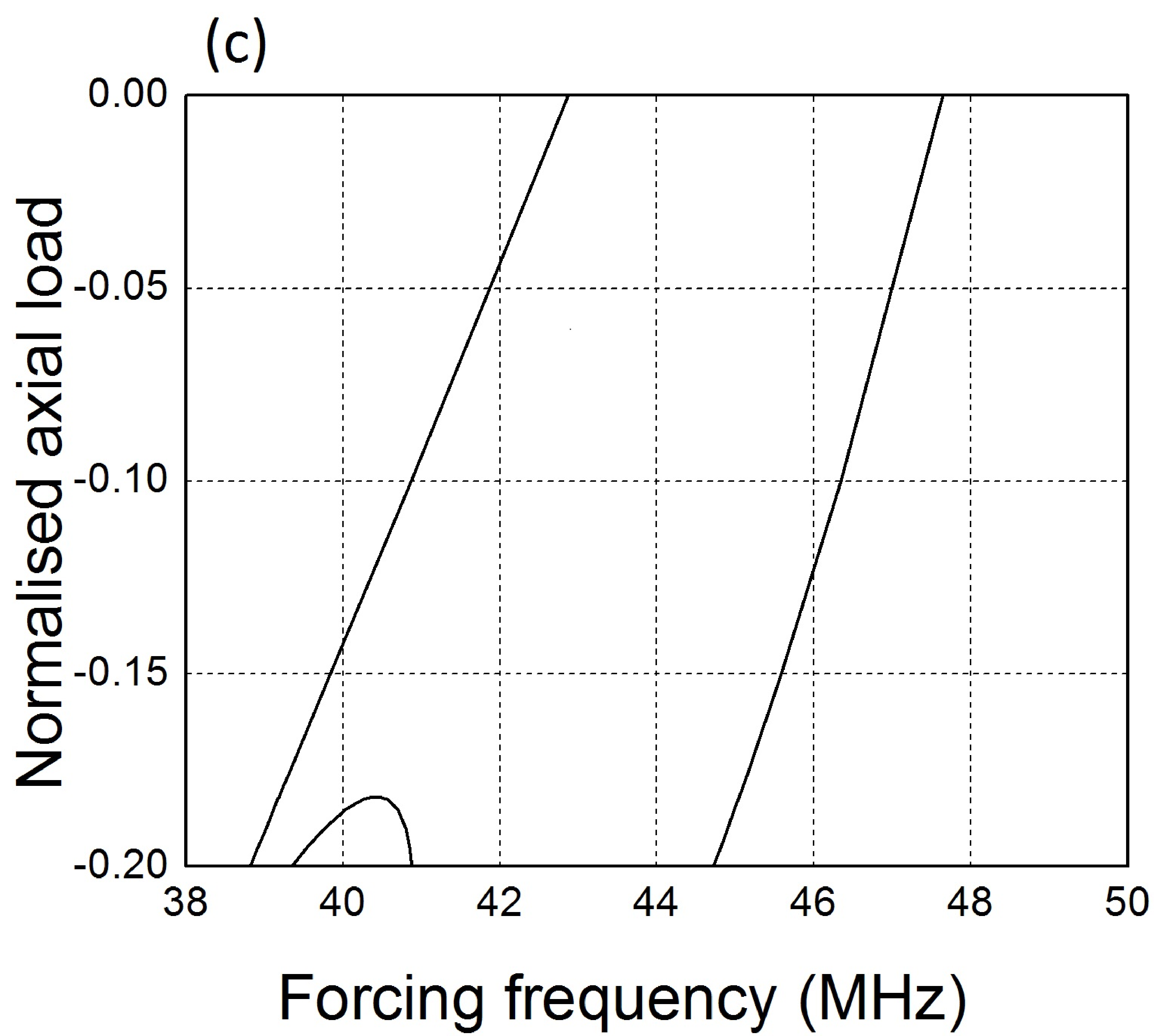}}
\hspace{0.5cm} 
\subfloat{\includegraphics[width=7.5cm]{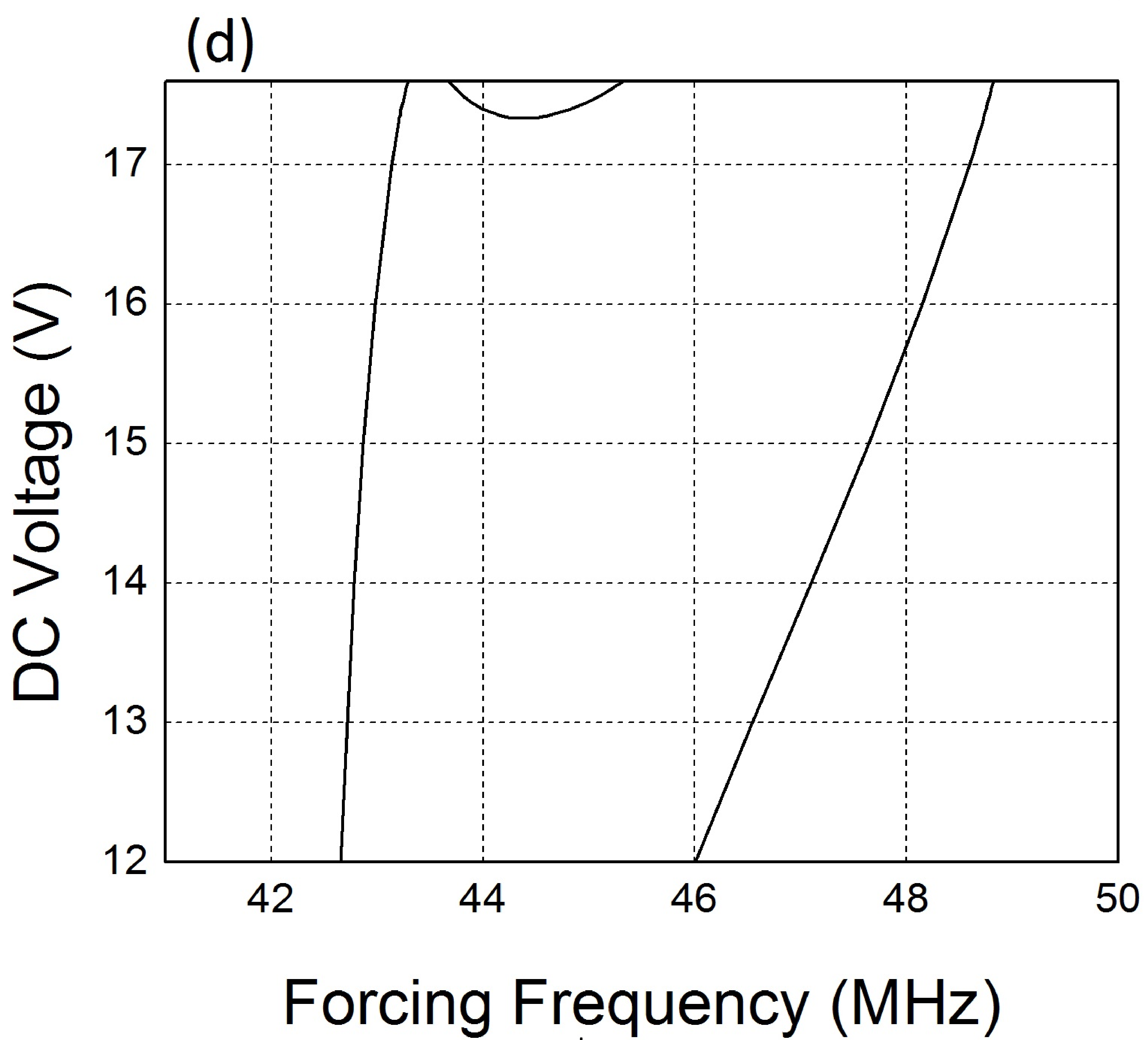}}
\caption{Branch bifurcation in ($V_{AC}- \omega_f$) plane for (a) $N = -0.2 N_{b1}$ and (b) $N = 0.2 N_{b1}$, respectively, and $V_{DC}= 15$ V is same in both diagrams. (c) Branch bifurcation in ($N_{nor}- \omega_f$) plane for $V_{DC}= 15$ V and $V_{AC}= 0.4$ V, $N_{nor}= N/N_{b1}$ is normalised axial load. (d) Branch bifurcation in ($V_{DC}- \omega_f$) plane for $N= 0$ and $V_{AC}= 0.4$ V. Here, $N_{b1} = 4 \pi^2$ is first Euler buckling load in non-dimensional form.}
\label{wfig.axial}
\end{figure*}

\begin{figure*}
\centering
\includegraphics[width=7.5cm]{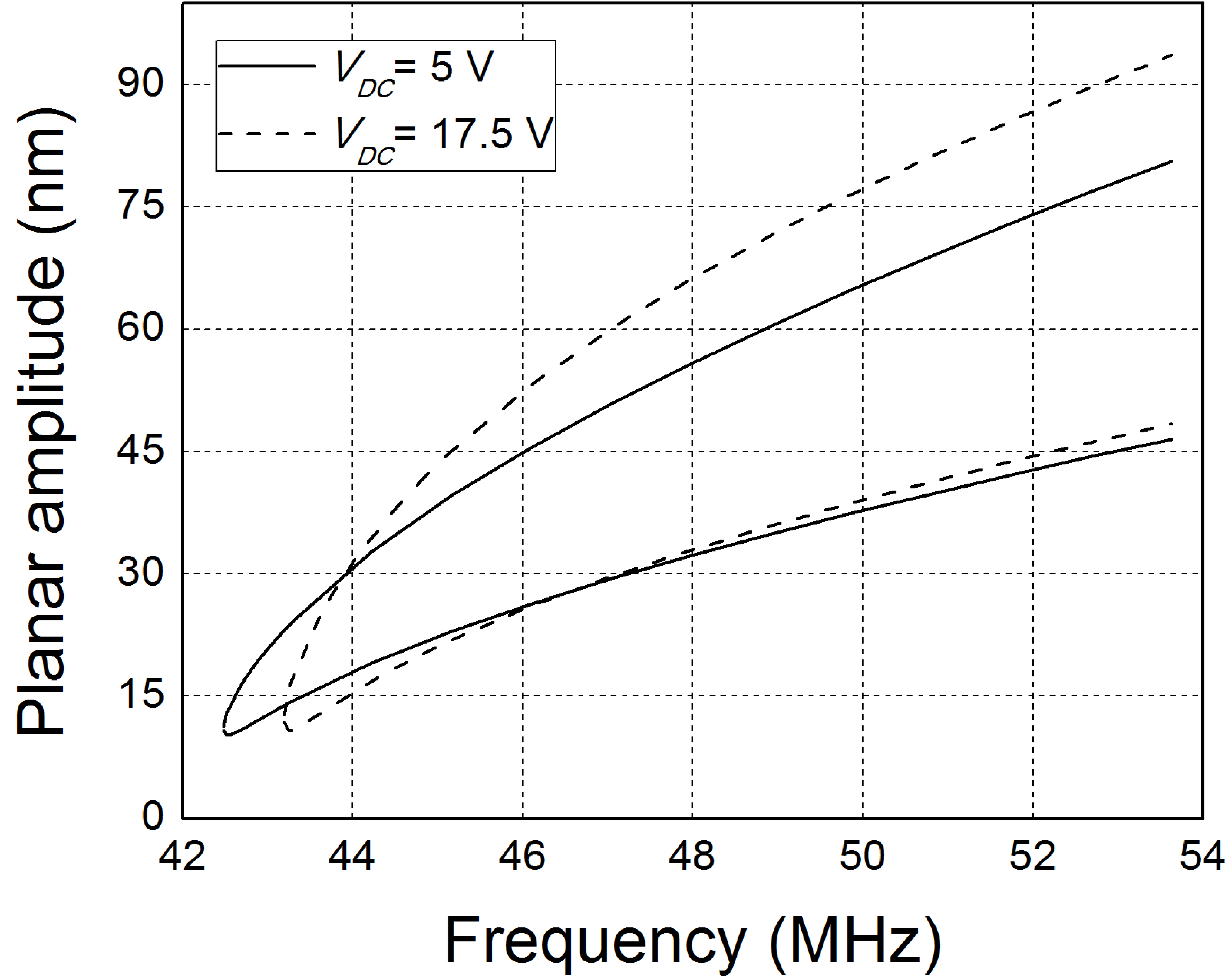}
\caption{Comparison of solutions of Eq. \eqref{weq.branch} for $V_{DC} = 5$ V and $V_{DC} = 17.5$ V.}
\label{wfig.instab1}
\end{figure*}

\begin{figure*}
\captionsetup[subfigure]{labelformat=empty}
\subfloat{\includegraphics[width=7.5cm]{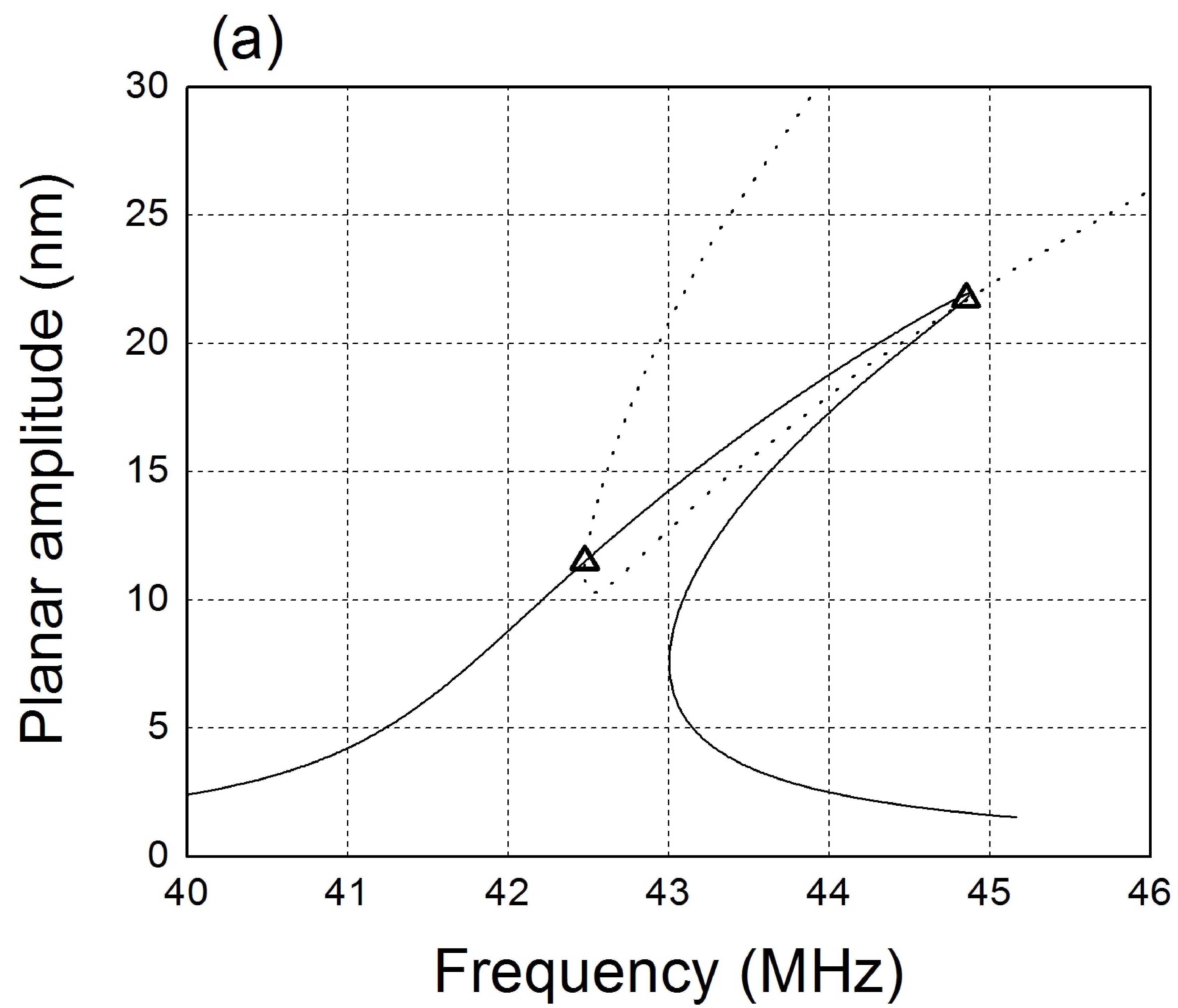}}
\hspace{0.5cm} 
\subfloat{\includegraphics[width=7.5cm]{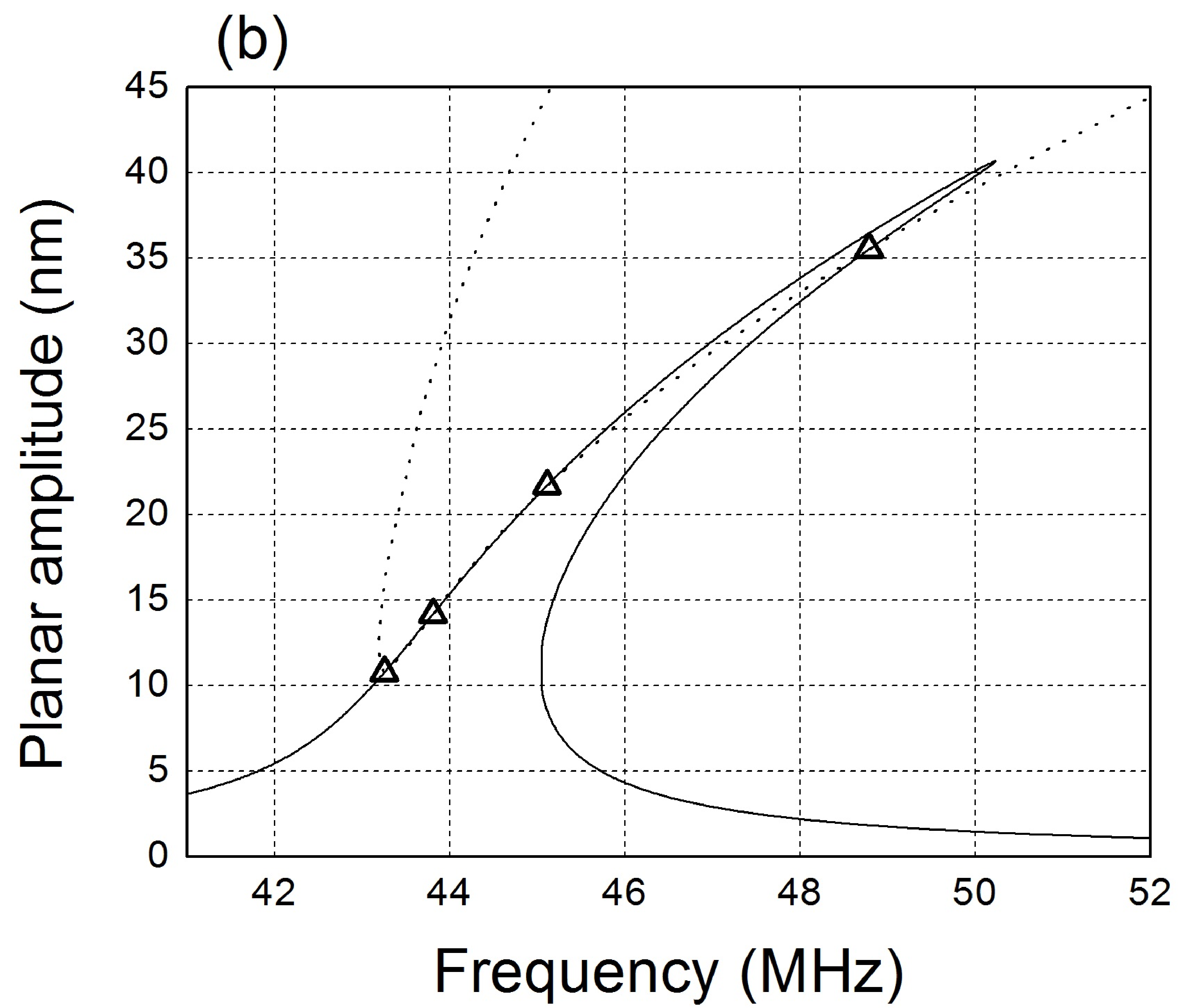}}
\caption{(a) Interaction of P-branch of planar resonance curve  with  solutions of Eq. \eqref{weq.branch} for (a) $V_{DC} = 5$ V and $V_{AC} = 0.7$ V and (b)  $V_{DC} = 17.5$ V and $V_{AC} = 0.4$ V.}
\label{wfig.instab2}
\end{figure*}

We have investigated bifurcation points of P-branch of resonance curves to study the effects of DC voltage on planar to whirling motion transition. In the absence of nonplanar motion, P-branch can be calculated by solving following algebraic equation, derived from Eq. \eqref{weq.averaged-eqn} by substituting $y_1 = r_u \cos (\theta_u)$, $y_2 = - r_u \sin (\theta_u)$, $y_3 = 0$, and $y_4 = 0$,
\begin{equation}\label{weq.planar-resonance}
\left( {p_0 } \right)^2  = r_u^2 \left( {\left( \mu_0  \right)^2  + \left( {\Omega _1  + \gamma _1 r_u^2 } \right)^2 } \right) \cdot
\end{equation}
For determining bifurcation points on P-branch of the resonance curves, Eq. \eqref{weq.averaged-eqn} has been re-written in the vector form as $\bf{ \dot y} = \textbf{G(y)}$, where $\textbf{y} = [y_1 \,\, y_2\,\, y_3\,\, y_4]^T$. The bifurcation on P-branch of resonance curves occurs when the determinant of the Jacobian matrix of $\textbf{G(y)}$ vanishes \cite{conley2008}. It is interesting to note that the Jacobian matrix, say $\textbf{B}$, is a four by four block diagonal matrix corresponding to the solutions of P-branch of the resonance curves. The saddle node bifurcation occurs when the determinant of first block matrix containing elements $B_{11}$, $B_{12}$, $B_{21}$, and $B_{22}$ vanishes. Hence the condition for saddle-node bifurcation is
\begin{equation} \label{weq.saddle}
\mu _0^2  + \left( {\Omega _1  + \gamma _1 r_u^2 } \right)\left( {\Omega _1  + 3\gamma _1 r_u^2 } \right) = 0 \cdot
\end{equation}
We have solved Eqs. \eqref{weq.planar-resonance} and \eqref{weq.saddle} simultaneously to obtain the saddle-node bifurcation points. Equation \eqref{weq.saddle} is only concerned with planar amplitude $r_u$ and is similar to the condition for saddle-node bifurcation of the averaged equation of Duffing oscillator \cite{murdock1991}. Figures \ref{wfig.bifs}(a) and \ref{wfig.bifs}(b) are the saddle-node bifurcation in AC voltage - forcing frequency ($V_{AC}-\omega_f$) plane for $V_{DC} = 5$ V and $V_{DC} = 17.5$ V respectively. To demonstrate the effectiveness of our analytical model, numerically obtained bifurcation points of the dynamical equations \eqref{weq.rom2} are correspondingly plotted in Fig. \ref{wfig.bifs}(a) and \ref{wfig.bifs}(b) as unfilled circles. One can observe from the figures that our analytical model is capable of obtaining the bifurcation points with reasonable accuracy. From Figs. \ref{wfig.bifs}(a) and \ref{wfig.bifs}(b), we can also observe that there are no saddle-node bifurcation below a threshold value of $V_{AC}$.\\
\indent Branch bifurcation corresponding to initiation of whirling motion occurs when the determinant of second block matrix, containing elements $B_{33}$, $B_{34}$, $B_{43}$, and $B_{44}$ vanishes \cite{lee1995}. This condition for branch bifurcation can be represented in the form of algebraic equation as
\begin{equation}\label{weq.branch}
\mu _0^2  + \left( {\Omega _2  + \gamma _3 r_u^2 } \right)\left( {\Omega _2  + (\gamma _3  - \gamma _4 )r_u^2 } \right) = 0 \cdot
\end{equation}
Figures \ref{wfig.bifw}(a), \ref{wfig.bifw}(b), \ref{wfig.bifw}(c), and \ref{wfig.bifw}(d) are the branch bifurcation in $V_{AC}-\omega_f$ plane for $V_{DC} = $ 5 V, 10 V, 15 V, and 17.5 V respectively. The diagrams have been obtained by simultaneously solving Eqs. \eqref{weq.planar-resonance} and \eqref{weq.branch}; these coupled algebraic equations can also be considered as an analytical criterion for initiation of whirling motion. Again, numerically obtained bifurcation points of Eq. \eqref{weq.rom2} are correspondingly plotted in Figs. \ref{wfig.bifw}(a)-\ref{wfig.bifw}(d) as unfilled triangles for demonstrating the effectiveness of the analytical model. There is overall qualitative agreement between the two solutions for the investigated range of $V_{DC}$; the error in the perturbation solution increases, for large magnitude of $V_{DC}$, as detuning parameters $\Omega_1$ and $\Omega_2$ increases. These figures demonstrate the change in qualitative behaviour of bifurcation pattern upon increasing DC voltage from 5 V to 17.50 V. As the magnitude of $V_{DC}$ increases, single local minima in bifurcation curves is modified to double minima. This results in a change in the number of branch bifurcation points on resonance curves, in case of $V_{DC} = 17.50$ V,  from two to four and restore to two as magnitude of $V_{AC}$ increases. However, when $V_{DC}$ is of lower magnitude, number of branch bifurcation points always remains two.  Furthermore, though, as can be seen from Fig. \ref{wfig.bifw}(a), for $V_{DC} = $ 5 V whirling motion is always initiated around $\hat \omega_f$ = 42.50 MHz, the initiation pattern is notably different for $V_{DC}$ = 17.5 V in Fig. \ref{wfig.bifw}(d). Specifically, whirling motion is initiated around $\hat \omega_f$ = 47 MHz at threshold magnitude of $V_{AC}$ = 0.29 V. Now with the increment of $V_{AC}$, the whirling initiation frequency decreases and finally becomes nearly invariable around 43.50 MHz for large magnitude of AC voltage.\\
\indent Till now, we have considered the nanowire to be free from axial load ($ N = 0$). We now study the effects of axial load on the whirling dynamics of the nanowire by observing change in bifurcation pattern with variation in axial load. In Figs. \ref{wfig.axial}(a) and \ref{wfig.axial}(b), respectively, two branch bifurcation curves corresponding to compressive axial load $N = -0.2 N_{b1}$ and tensile axial load $N = 0.2 N_{b1}$ are plotted for $V_{DC} = 15$ V. Here, $N_{b1}$ is non-dimensional first Euler-Buckling load of a doubly-clamped beam and its value is $4\pi^2$. The qualitatively similar nature of change in branch bifurcation pattern can be observed in Figs. \ref{wfig.axial}(a) and \ref{wfig.axial}(b) with decrement of axial load (reduction in tensile axial load or rise in compressive axial load), as we have observed with increment of DC voltages in Figs. \ref{wfig.bifw}(a)-\ref{wfig.bifw}(d). This observation is further reflected in Figs. \ref{wfig.axial}(c) and \ref{wfig.axial}(d). Figure \ref{wfig.axial}(c) shows branch bifurcation pattern in ($N_{nor}- \omega_f$) plane for $V_{DC}= 15$ V and $V_{AC}= 0.4$ V, $N_{nor}= N/N_{b1}$ is normalised axial load. Similarly, Fig. \ref{wfig.axial}(d) shows branch bifurcation pattern in ($V_{DC}- \omega_f$) plane for $N= 0$ and $V_{AC}= 0.4$ V. We have discussed in Section 3 that quadratic nonlinearities are strongly dependent on static deflection of the nanowire. With increment of DC voltage, static deflection increases and also coefficients of quadrantal nonlinearities increases (refer Table \ref{wtab.nonlinear}), which results in qualitative change in bifurcation pattern. Similar effects on quadratic nonlinearities will also be observed with decrement of axial load for same magnitude of DC voltage, and this is the reason for similarity in change in bifurcation pattern with decrement of axial load and increment of DC voltage.\\
\indent We have further examined the dependence of nature of branch bifurcation on the magnitude of DC voltage by studying the solution of Eqs. \eqref{weq.planar-resonance} and \eqref{weq.branch} in amplitude-frequency ($r_u - \omega_f$) plane. Solution of Eq. \eqref{weq.planar-resonance} provide the P-branch of resonance curve in $r_u - \omega_f$ plane, whereas the solution of Eq. \eqref{weq.branch} provides the condition for branch bifurcation on P-branch of resonance curves. Upon comparing the solutions of Eq. \eqref{weq.branch} for $V_{DC} = 5$ V and $V_{DC} = 17.5$ V in Fig. \ref{wfig.instab1}, one can observe that qualitative nature of both the curves is same. However, the interaction of the P-branch of resonance curve with the curve for bifurcation condition can qualitatively change with increase in the magnitude of DC voltage. To clarify this, the P-branch of the planar resonance curve obtained from Eq. \eqref{weq.planar-resonance} is plotted along with the bifurcation condition resulting from Eq. \eqref{weq.branch} for $V_{DC} = 5 $ V and $V_{AC} = 0.70$ V (Fig. \ref{wfig.instab2}(a)) and  for $V_{DC} = 17.5 $ V and $V_{AC} = 0.40$ V (\ref{wfig.instab2}(b)). We can observe from these figures that though the resonance and bifurcation curves have two intersection points for $V_{DC} = $ 5 V, the number of intersection points jump to four when $V_{DC}$ = 17.5 V. Hence, we conclude that variation in the magnitude of DC voltage leads to a qualitative change in the nature of planar and nonplanar resonance curves.

\section{Conclusion}
We have modelled a nanowire oscillator as a two-degree of freedom system using Galerkin based reduced order modelling technique. The dynamical equation of motion is corresponding to nonlinear asymmetric coupled oscillator problem, and has been solved using both numerical technique and analytical technique second-order averaging method. We have systematically investigated the effect of DC voltage and axial load on planar to whirling motion transition and observed that qualitative nature of resonance curves can be tuned by controlling DC and AC voltages. We have also observed that magnitude of axial load in nanowire may have significant effect on qualitative nature of resonance curves. Using averaging solution, physical insights have been gained to understand the relation between magnitude of DC voltage and whirling initiation pattern of the nanowire oscillator. A simple analytical condition in form of coupled algebraic equations has also been provided to detect the initiation of whirling motion in nanowire oscillators. The analytical condition will be a helpful tool for practicing engineers and scientists during designing nano-oscillators to quantify the critical magnitude of DC and AC voltages to avoid/initiate whirling motion.
\section*{Acknowledgments}
Financial support from Department of Science and Technology, Government of India (Grant No. - SR/FTP/ETA-031/2009) is gratefully acknowledged.
\bibliographystyle{unsrt}
\bibliography{manuscript}
\end{document}